\documentclass[aps,prd,reprint,groupedaddress,eprint,preprintnumbers]{revtex4-2}

\makeatletter
\newif\ifpreprintoption
\@ifclasswith{revtex4-2}{preprint}{\preprintoptiontrue}{\preprintoptionfalse}
\makeatother
\newcommand{\ifpreprint}[2]{\ifpreprintoption #1\else #2\fi}
\usepackage{scalerel}
\usepackage{tikz}
\usetikzlibrary{svg.path}

\definecolor{orcidlogocol}{HTML}{A6CE39}
\tikzset{
  orcidlogo/.pic={
    \fill[orcidlogocol] svg{M256,128c0,70.7-57.3,128-128,128C57.3,256,0,198.7,0,128C0,57.3,57.3,0,128,0C198.7,0,256,57.3,256,128z};
    \fill[white] svg{M86.3,186.2H70.9V79.1h15.4v48.4V186.2z}
                 svg{M108.9,79.1h41.6c39.6,0,57,28.3,57,53.6c0,27.5-21.5,53.6-56.8,53.6h-41.8V79.1z M124.3,172.4h24.5c34.9,0,42.9-26.5,42.9-39.7c0-21.5-13.7-39.7-43.7-39.7h-23.7V172.4z}
                 svg{M88.7,56.8c0,5.5-4.5,10.1-10.1,10.1c-5.6,0-10.1-4.6-10.1-10.1c0-5.6,4.5-10.1,10.1-10.1C84.2,46.7,88.7,51.3,88.7,56.8z};
  }
}

\newcommand\orcidicon[1]{\href{https://orcid.org/#1}{\mbox{\scalerel*{
\begin{tikzpicture}[yscale=-1,transform shape]
\pic{orcidlogo};
\end{tikzpicture}
}{0}}}}

\usepackage{graphicx}
\usepackage[utf8]{inputenc} 
\usepackage[T1]{fontenc}    
\usepackage{xcolor}
\definecolor{darkblue}{RGB}{46,48,147}
\usepackage[colorlinks=true,
            linkcolor=darkblue,
            urlcolor=darkblue,
            citecolor=darkblue]{hyperref}
\usepackage{url}            
\usepackage{booktabs}       
\usepackage{amsfonts}       
\usepackage{nicefrac}       
\usepackage{microtype}      
\usepackage{amsmath}
\usepackage{multirow}
\usepackage{xspace}
\usepackage{amssymb}

\providecommand{\mSD}{\ensuremath{m_\mathrm{SD}}\xspace} 
\providecommand{\PH}{\ensuremath{H}\xspace} 
\providecommand{\PZ}{\ensuremath{Z}\xspace} 
\providecommand{\PW}{\ensuremath{W}\xspace} 
\providecommand{\PV}{\ensuremath{V}\xspace} 
\providecommand{\Pg}{\ensuremath{g}\xspace} 
\providecommand{\PQc}{\ensuremath{c}\xspace} 
\providecommand{\PQb}{\ensuremath{b}\xspace} 
\providecommand{\PAQb}{\ensuremath{\overline{b}}\xspace} 
\providecommand{\PQt}{\ensuremath{t}\xspace} 
\providecommand{\PAQt}{\ensuremath{\overline{t}}\xspace} 
\providecommand{\PQq}{\ensuremath{q}\xspace} 
\newcommand{\bbbar}{\ensuremath{\PQb\PAQb}\xspace}
\newcommand{\ttbar}{\ensuremath{\PQt\PAQt}\xspace}

\newcommand{\pt}{\ensuremath{p_{\mathrm{T}}}\xspace}
\newcommand{\kt}{\ensuremath{k_{\mathrm{T}}}\xspace}
\newcommand{\PYTHIA} {{\textsc{pythia}}\xspace}
\newcommand{\GEANTfour} {{\textsc{Geant4}}\xspace}
\newcommand{\MADGRAPH} {{\textsc{MADGRAPH}}\xspace}
\newcommand{\MCATNLO} {{\textsc{MCATNLO}}\xspace}
\newcommand{\MGvATNLO}{\MADGRAPH{}5\_a\MCATNLO\xspace}
\newcommand{\GeV}{\ensuremath{\,\text{Ge\hspace{-.08em}V}}\xspace}
\newcommand{\TeV}{\ensuremath{\,\text{Te\hspace{-.08em}V}}\xspace}

\newcommand{\ptvecmiss}{\ensuremath{{\vec p}_{\mathrm{T}}^{\kern1pt\text{miss}}}\xspace}
\graphicspath{{figures/}}
\received{10 December 2019}
\accepted{13 June 2020}
\published{28 July 2020}

\begin{document}

\preprint{FERMILAB-PUB-19-492-CMS-E}

\title{Interaction networks for the identification of boosted $\boldsymbol{\PH\to\bbbar}$ decays}

\author{Eric A. Moreno\,\orcidicon{0000-0001-5666-3637}}
\author{Thong Q. Nguyen\,\orcidicon{0000-0003-3954-5131}}
\author{Jean-Roch Vlimant\,\orcidicon{0000-0002-9705-101X}}
\author{Olmo~Cerri\,\orcidicon{0000-0002-2191-0666}}
\author{Harvey B. Newman\,\orcidicon{0000-0003-0964-1480}}
\author{Avikar Periwal}
\author{Maria Spiropulu\,\orcidicon{0000-0001-8172-7081}}
\affiliation{California Institute of Technology, Pasadena, California 91125, USA}
\author{Javier M. Duarte\,\orcidicon{0000-0002-5076-7096}}
\email{jduarte@ucsd.edu}
\affiliation{University of California San Diego, La Jolla, California 92093, USA}
\affiliation{Fermi National Accelerator Laboratory, Batavia, Illinois 60510, USA}
\author{Maurizio Pierini\,\orcidicon{0000-0003-1939-4268}}
\affiliation{European Organization for Nuclear Research, 1211 Geneva 23, Switzerland\vspace{\baselineskip}}

\begin{abstract}
  We develop an algorithm based on an interaction network to identify high-transverse-momentum Higgs bosons decaying to bottom quark-antiquark pairs and distinguish them from ordinary jets that reflect the configurations of quarks and gluons at short distances.
  The algorithm's inputs are features of the reconstructed charged particles in a jet and the secondary vertices associated with them.
  Describing the jet shower as a combination of particle-to-particle and particle-to-vertex interactions, the model is trained to learn a jet representation on which the classification problem is optimized.
  The algorithm is trained on simulated samples of realistic LHC collisions, released by the CMS Collaboration on the CERN Open Data Portal.
  The interaction network achieves a drastic improvement in the identification performance with respect to state-of-the-art algorithms.\\\\
  DOI: \href{https://doi.org/10.1103/PhysRevD.102.012010}{10.1103/PhysRevD.102.012010}
 
\end{abstract}

\maketitle

\section{Introduction}
\label{sec:intro}

Jets are collimated showers of hadrons that reflect the configurations of quarks and gluons produced at particle colliders.
Each shower, consisting of quarks and gluons emitted by the primary particle, results in an approximately cone-shaped spray of hadrons, which are then observed in particle detectors.
Jet identification, or \emph{tagging}, algorithms are designed to identify the nature of the primary particle that initiates a shower by studying the collective features of the hadrons inside the jet.

\begin{figure}[htpb!]
    \centering
    \includegraphics[width=\columnwidth]{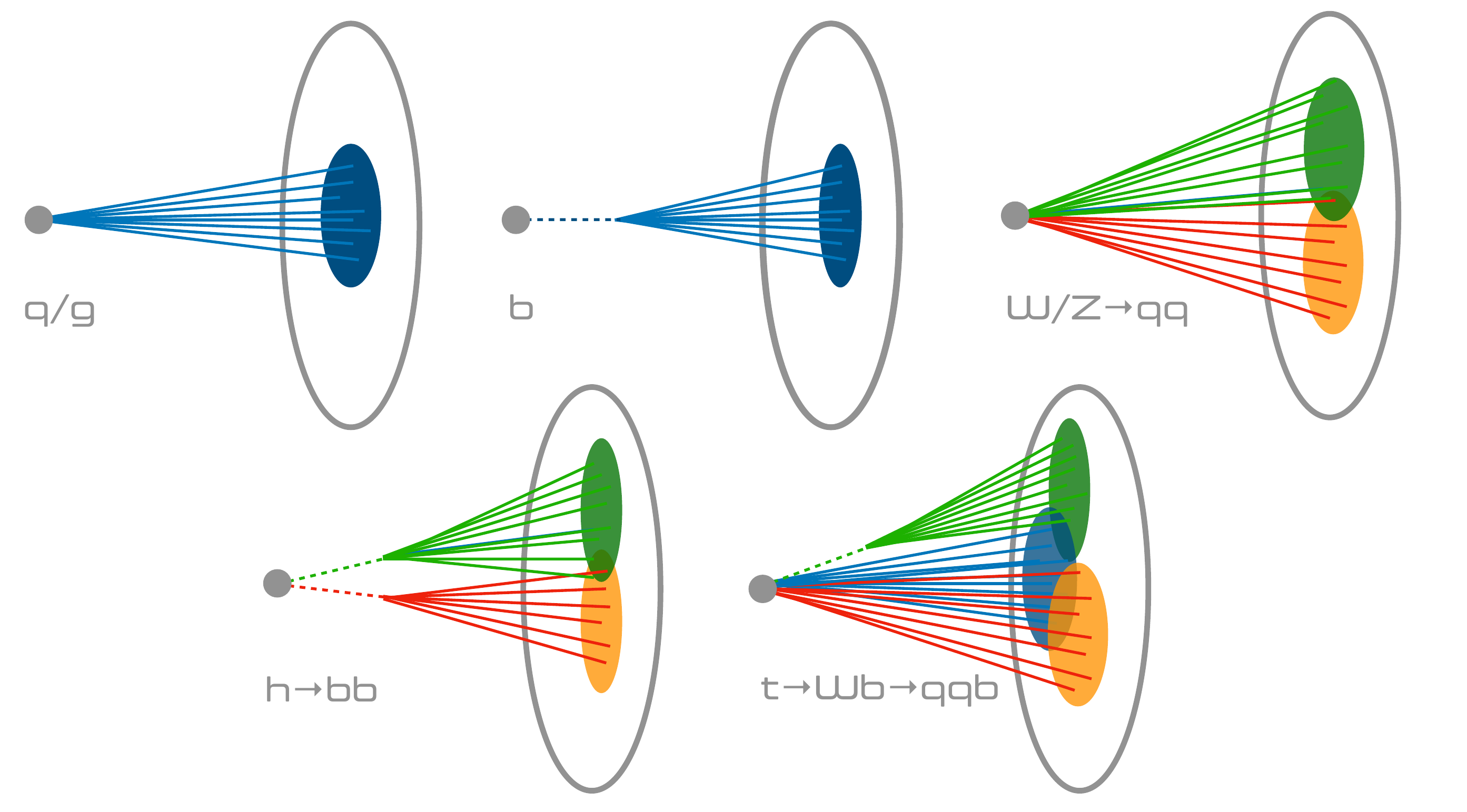}
    \caption{Pictorial representation of ordinary quark and gluon jets (top left), $\PQb$ jets (top center), and boosted-jet topologies, emerging from high-$\pt$ $\PW$ and $\PZ$ bosons (top right), Higgs bosons (bottom left), and top quarks (bottom right) decaying to all-quark final states.\label{fig:boostedJetCartoon}}
\end{figure}

Traditionally, jet tagging was limited to light-flavor quarks ($\PQq$), gluons ($\Pg$), or $\PQb$ quarks.
At the CERN Large Hadron Collider (LHC), jet tagging becomes a more complex task as new jet topologies are accessible (see Fig.~\ref{fig:boostedJetCartoon}).
Due to the large center-of-mass energy available in LHC collisions, heavy particles, such as $\PW$, $\PZ$, and Higgs bosons ($\PH$) and top quarks ($\PQt$), may be produced with large transverse momentum ($\pt$).
These particles can decay to all-quark final states.
Due to the large $\pt$ of the original particle, these quarks are produced within a small solid angle.
The overlapping showers produced by these quarks may be reconstructed as a single massive jet.
As shown in Fig.~\ref{fig:boostedJetCartoon}, the presence of $\PQb$ quarks in the jet gives rise to unique experimental signatures.
In particular, $\PQb$ hadrons are characterized by a lifetime of approximately $1.5$~ps, which results in a detectable displacement between the proton-collision point and the point where the $\PQb$ hadron decays.

The identification of jets from heavy resonances relies on \emph{jet substructure} techniques, designed to quantify the number of clusters of energetic particles, or \emph{prongs}, inside the jet.
The study of jet substructure was pioneered in the 1990s and early 2000s~\cite{Seymour:1991cb,Seymour:1993mx,Seymour:1994by,Butterworth:2002tt}, but interest skyrocketed after its proposed application to reconstruct Higgs bosons when produced in association with a vector boson~\cite{Butterworth:2008iy}.
Extensive reviews of these techniques are provided in Refs.~\cite{Larkoski:2017jix,Marzani:2019hun}.
Additional discrimination is provided by the reconstructed jet mass, usually computed after a jet grooming algorithm.
A review of the techniques used to reconstruct jets and their substructure at the LHC experiments can be found in Ref.~\cite{Asquith:2018igt}.
The jet mass plays a special role in physics analyses exploiting jet substructure, as described for instance in Ref.~\cite{Sirunyan:2019jbg}.
The jet mass distribution is typically used to separate jets from boosted heavy particles, characterized by a peaking distribution, from the smoothly falling background, due to ordinary quark and gluon jets.
For certain applications, it is desirable to avoid any distortion of the jet mass distribution when applying a jet-tagging selection.

Due to its lifetime, the presence of a $\PQb$ hadron inside of a jet typically results in a reconstructed secondary vertex (SV) that is displaced from the primary vertex (PV).
Modern particle detectors are equipped with a vertex detector that can accurately determine SV positions and their separation from the PV, even in a dense environment like a high-$\pt$ jet.
This feature is particularly important for tagging a Higgs boson decaying to a bottom quark-antiquark pair ($\PH\to\bbbar$) because all of the relevant jet constituents originate from two displaced vertices.

Recently, several approaches based on deep learning (DL) have been proposed to optimize jet tagging algorithms (see Sec~\ref{sec:relatedWorks}), both using expert features with dense layers or raw data representations (e.g., images or lists of particle properties) with more complex architectures.
For instance, the LHC collaborations and other researchers have investigated the optimal way to combine substructure, tracking, and vertexing information to enhance the tagging efficiency for high-$\pt$ $\PH\to\bbbar$ decays~\cite{Sirunyan:2017ezt, CMS-Hbb-baseline, CMS-DP-2017-049,Sirunyan:2020lcu,Aad:2019uoz,Lin:2018cin}.
This is an important task in particle physics because measurements of high-$\pt$ $\PH\to\bbbar$ decays may help resolve the loop-induced and tree-level contributions to the gluon fusion process, providing an complementary approach to study the $\PQt$ Yukawa beyond the $\ttbar\PH$ process~\cite{Sirunyan:2017dgc,Sirunyan:2018sgc,Grojean:2013nya,Becker:2669113}.
These measurements are also sensitive probes for physics beyond the standard model~\cite{ATLAS:2018hzj,Sirunyan:2018sgc,Grojean:2013nya,Dawson:2015gka,Schlaffer:2014osa,Grazzini:2017szg,Grazzini:2016paz,Bishara:2016jga,Li:2019pag}.
Finally, improving these measurements is important for measuring the Higgs boson self-coupling through the production of $\PH\PH\to\bbbar\bbbar$~\cite{Amacker:2020bmn,Dainese:2703572,Kling:2016lay,Grazzini:2018bsd}.

While existing DL approaches have been successfully applied to jet tagging, particle jets involve multiple entities with complex interactions that are not easily encoded as images or lists.
Graphs provide a natural representation for such relational information.
Traditional machine learning methods use feature engineering and preprocessing to learn from these graphs, which can be time consuming and costly, and may miss important features present in the data.
Graph representation learning, including graph convolution networks~\cite{niepert2016learning, kipf2016semi,DBLP:journals/corr/QiSMG16,DBLP:journals/corr/abs-1801-07829} and graph generative models~\cite{grover2018graphite,you2018graphrnn}, leverages DL to learn directly from graph-structured data.
In contrast to other DL methods, graph representation learning can (1) handle irregular grids with non-Euclidean geometry~\cite{bruna2013spectral}, (2) encode physics knowledge via graph construction~\cite{zheng2018unsupervised}, and (3) introduce relational inductive bias into data-driven learning systems~\cite{battaglia2018relational}.
For example, while convolutional neural networks (CNNs) are powerful classifiers that work extremely well for data represented on a grid~\cite{cnn, resnet}, geometric DL algorithms, such as graph neural networks (GNNs)~\cite{gnn1, gnn2}, are applicable even without an underlying grid structure.
Because the data in many scientific domains are not Euclidean, GNNs emerge as a more natural choice.

In this work, we propose to identify $\PH\to\bbbar$ jets with an interaction network (IN), a type of graph network.
In Ref.~\cite{interactionnetwork}, INs were introduced to describe complex physical systems and predict their evolution after a certain amount of time.
This was achieved by constructing graph networks to learn the interactions between the physical objects, represented as the nodes of the graph.
Just as noted jet substructure variables like $N^{\beta=1}_2$ and $D^{\beta=1}_2$ compute 2-point energy correlation functions between jet constituents to quantify the number of prongs in a jet~\cite{Larkoski:2013eya,Moult:2016cvt}, we posit that the ability of INs to learn complex pairwise relationships aids in identifying the patterns present in $\PH\to\bbbar$ decays.
Moreover, Ref.~\cite{jedi-net} showed that the learned features of an IN correlate with known jet substructure variables.
It was further demonstrated that the IN architecture outperformed other deep neural networks (DNNs), such as dense, convolutional, and recurrent networks, for a jet-substructure classification task.
However, this study was limited because the simulation considered was not fully realistic.

In this paper, we demonstrate that an interaction network with an extended feature representation outperforms state of the art methods for $\PH\to\bbbar$ tagging with \GEANTfour-based~\cite{GEANT4} realistic simulation, while relying on less parameters.
In particular, we investigate the use of INs to learn a collective representation of the tracking, vertexing, and substructure properties of the jet and employ this optimized representation to enhance the tagging efficiency.
By placing charged particles and secondary vertices on a graph, the network can learn a representation of each particle-to-particle and particle-to-vertex \emph{interaction}, and exploit this information to categorize a given jet as signal ($\PH\to\bbbar$) or background (QCD).

The study is carried out using a sample of fully simulated LHC collision events, released by the CMS Collaboration on the CERN Open Data portal~\cite{CERNOpenDataPortal}.
Previously, many machine learning studies were limited to studies based on generator-level physics with simple detector emulation.
The released CMS full-simulation samples allow for a more in depth and realistic study of the efficacy of machine learning methods on high-energy physics experiments.
We compare the performance to several different algorithms that we trained with open simulation for $\PH\to\bbbar$ tagging based on the architecture of the deep double-$\PQb$ (DDB) tagger created by the CMS Collaboration~\cite{CMS-Hbb-baseline}.

The IN and DDB taggers only rely on information related to charged particles, which (unlike neutral particles) can be traced back to their point of origin: the PV of the high-$\pt$ collision, any SV generated in the collision, or additional PVs originating from simultaneous proton-proton interactions (pileup).
This choice makes the algorithm particularly robust against the large pileup contamination expected in future LHC runs since this contamination can be removed via so-called charged hadron subtraction (CHS)~\cite{Kirschenmann:2014dla}.
For the IN tagger, we consider an extended representation of each charged particle (secondary vertex), with 22 (12) additional features with respect to the nominal DDB tagger (as discussed in Sec.~\ref{sec:data}).
To enable a fair comparison between network architectures, we also report results for an extended variant of the DDB tagger, the deep double-\PQb+ (DDB+) tagger, which consumes the same information as the IN tagger.

This paper is structured as follows: we discuss related work in Sec.~\ref{sec:relatedWorks}.
Section~\ref{sec:data} gives a brief description of the datasets used.
Sections~\ref{sec:in}~and~\ref{sec:decorrelation} describe the IN architecture and the algorithms used to decorrelate its score from the jet mass distribution.
Section~\ref{sec:ddb} describes our reconstruction and training of the DDB and DDB+ algorithms.
Results are presented in Sec.~\ref{sec:results} and conclusions are given in Sec.~\ref{sec:conclusions}.

\section{Related work}
\label{sec:relatedWorks}

The use of DNNs has recently found a great deal of success in particle physics~\cite{Larkoski:2017jix,Guest:2018yhq}, especially jet tagging.
Driving this innovation are increasingly complex architectures that are tailored to particular domains, including CNNs~\cite{lecun1995convolutional, lawrence1997face, krizhevsky2012imagenet}, which are well suited to computer vision, and recurrent neural networks (RNNs)~\cite{williams1989learning, graves2013speech} like long short-term memory units (LSTMs)~\cite{hochreiter1997long} and gated recurrent units (GRUs)~\cite{chung2014empirical}, which are appropriate for natural language processing.
Several classification algorithms have been studied in the context of jet tagging at the LHC using CNNs~\cite{deOliveira:2015xxd,Macaluso:2018tck,Kasieczka:2017nvn,Komiske:2016rsd,Baldi:2016fql} and physics-inspired DNN models~\cite{Datta:2017lxt,Butter:2017cot,Komiske:2017aww,Baldi:2016fql}.
Recurrent and recursive layers have been used to define jet classifiers starting from a list of reconstructed particle momenta~\cite{RecursiveJets,Egan:2017ojy,Cheng:2017rdo,Guest:2016iqz}.
Recently, several different approaches, applied to the specific case of $\PQt$ jet identification have been compared~\cite{Kasieczka:2019dbj} on a public $\PQt$ jet tagging dataset~\cite{kasieczka_gregor_2019_2603256}.
This study found ParticleNet~\cite{Qu:2019gqs}, a GNN based on the dynamic graph CNN~\cite{DBLP:journals/corr/abs-1801-07829} to be the best performing for that task.
In Ref.~\cite{jedi-net}, it was shown that the area under the receiver operating characteristic (ROC) curve (AUC), accuracy, and background rejection at a 30\% true positive rate (TPR) of a simple IN architecture trained with the same dataset is within 1\%, 0.5\%, and 40\% of those of ParticleNet, while using 70\% fewer parameters.
Unsupervised, semisupervised, and weakly supervised methods have also been proposed, mainly to tag $\PQt$ jets or jets coming from postulated new particles~\cite{Heimel:2018mkt,Farina:2018fyg,Dillon:2019cqt,Collins:2018epr,Collins:2019jip,Dery:2017fap,Metodiev:2017vrx,Komiske:2018oaa,Cohen:2017exh,Nachman:2020lpy,Andreassen:2020nkr}.
Finally, others have also explored the CMS open data and simulation to study jet properties and jet classification algorithms in a realistic setting~\cite{Tripathee:2017ybi,Larkoski:2017bvj,Andrews:2018nwy,Andrews:2019faz,Komiske:2019fks,Komiske:2019jim}.

For the task of identifying $\PH\to\bbbar$ specifically, several machine learning approaches have been applied.
In generator-level studies, Ref.~\cite{Lin:2018cin} uses images, representing both the $\PH$ candidate jet and the full event, as inputs to a CNN.
In conditions more closely resembling real data, the CMS Collaboration created a boosted decision tree based on expert chosen features to identify the presence of two $\PQb$ hadrons within a single anti-$\kt$~\cite{Cacciari:2008gp, Cacciari:2011ma} $R=0.8$ jet (AK8 jet)~\cite{Sirunyan:2017ezt}.
This approach was extended using a deep neural network and additional particle-level and vertex level information, the DDB tagger~\cite{CMS-Hbb-baseline}.
Other more generic CMS algorithms, also based on deep neural networks and known as the boosted event shape tagger (BEST) and the DeepAK8 tagger, were created to classify the decays of multiple heavy resonances, including $\PH$, $\PZ$, $\PW$, and $\PQt$~\cite{Sirunyan:2020lcu}.
The ATLAS collaboration has also designed an algorithm to identify two $\PQb$ hadrons within an anti-$\kt$ $R=1$ jet using $\PQb$ tagging of track-based subjets~\cite{Aad:2019uoz}.
For the task of $\PH\to\bbbar$ identification, the CMS DDB tagger, DeepAK8 algorithm, and the ATLAS tagger achieve similar state-of-the-art performance.

Graph networks~\cite{graphjettagging,Qu:2019gqs,jedi-net,Kasieczka:2019dbj} and the related particle flow networks~\cite{Komiske:2018cqr} have recently been used for other kinds of jet tagging, matching or exceeding the performances of other DL approaches, for event classification~\cite{Abdughani:2018wrw,Choma2018GraphNN}, for charged particle tracking in a silicon detector~\cite{GraphTracking,ExaTrkX}, for mitigation of the effects pileup~\cite{Martinez:2018fwc}, and for particle reconstruction in irregular calorimeters~\cite{Qasim:2019otl,Kieseler:2020wcq,Gray:2020mcm,ExaTrkX} and the IceCube experiment~\cite{Choma2018GraphNN}.

While applying GNNs is natural for particle physics data, one issue we confront in this paper is how to deal with heterogeneous hierarchical data, i.e. data composed of different sets of elements with different numbers and types of features.
The primary original contributions of this paper are (1) designing an IN with data comprising a heterogeneous graph with two types of graph nodes: particles and SVs), (2) demonstrating that an IN achieves competetive performance on public, realistic simulation for the task of $\PH\to\bbbar$ tagging with fewer trainable parameters in a way that is robust to the effects of pileup, and (3) comparing and evaluating mass decorrelation methods.

\section{Data samples}
\label{sec:data}

The CMS open data and simulation are available from the CERN Open Data Portal~\cite{CERNOpenDataPortal}, including releases of 2010, 2011, and 2012 CMS collision data as well as 2011, 2012, and 2016 CMS simulated data.

Samples of $\PH\to\bbbar$ jets are available from simulated events containing Randall-Sundrum gravitons~\cite{Randall} decaying to two Higgs bosons, which subsequently decay to $\bbbar$ pairs.
The event generation was done by the CMS Collaboration with \MGvATNLO~2.2.2 at leading order, with graviton masses ranging between 0.6 and 4.5\TeV.
Generation of this process enables better sampling of events with large Higgs boson $\pt$.
The main source of background originates from multijet events.
The background dataset was generated with \PYTHIA~8.205~\cite{Sjostrand:2014zea} in different bins of the average $\pt$ of the final-state partons ($\hat{p}_\mathrm{T}$).
The parton showering and hadronization was performed with \PYTHIA~8.205~\cite{Sjostrand:2014zea}, using the CMS underlying event tune CUETP8M1~\cite{Khachatryan:2015pea} and the NNPDF~2.3~\cite{NNPDF2} parton distribution functions.
Pileup interactions are modeled by overlaying each simulated event with additional minimum bias collisions, also generated with \PYTHIA~8.205.
The CMS detector response is modeled by \GEANTfour~\cite{GEANT4}.

The outcome of the default CMS reconstruction workflow is provided in the open simulation~\cite{opendata}.
In particular, particle candidates are reconstructed using the particle-flow (PF) algorithm~\cite{CMS-PRF-14-001}.
Charged particles from pileup interactions are removed using the CHS algorithm.
Jets are clustered from the remaining reconstructed particles using the anti-$\kt$ algorithm~\cite{Cacciari:2008gp, Cacciari:2011ma} with a jet-size parameter $R=0.8$.
The standard CMS jet energy corrections are applied to the jets.
In order to remove soft, wide-angle radiation from the jet, the soft-drop (SD) algorithm~\cite{Dasgupta:2013ihk,Butterworth:2008iy} is applied, with  angular exponent $\beta = 0$, soft cutoff threshold $z_{\mathrm{cut}} < 0.1$, and characteristic radius $R_{0} = 0.8$~\cite{Larkoski:2014wba}.
The SD mass ($\mSD$) is then computed from the four-momenta of the remaining constituents.

A signal $\PH\to\bbbar$ jet is defined as a jet geometrically matched to the generator-level Higgs boson and both $\PQb$ quark daughters.
Jets from QCD multijet events are used to define a sample of fake $\PH\to\bbbar$ candidates.

The dataset is reduced by requiring the AK8 jets to have $300 < \pt < 2400 \GeV$, $|\eta| < 2.4$, and $40 < \mSD < 200 \GeV$.
After this reduction, the dataset consists of 3.9 million $\PH\to\bbbar$ jets and 1.9 million inclusive QCD jets.
Charged particles are required to have $\pt > 0.95 \GeV$ and reconstructed secondary vertices (SVs) are associated with the AK8 jet using $\Delta R = \sqrt{\Delta\phi^2+\Delta\eta^2} < 0.8$.
The dataset is divided into blocks of features, referring to different objects.
Different blocks are used as input by the models described in the rest of the paper.

The IN uses 30 features related to charged particles (see Table~\ref{tab:track_features} in Appendix~\ref{sec:features}).
The IN also uses 14 SV features listed in Table~\ref{tab:sv_features}.
The DDB tagger~\cite{CMS-Hbb-baseline} uses a subset of the above features (8 features for each particle and 2 features for each SV), chosen to minimize the correlation with the jet mass.
In addition, the DDB tagger uses 27 high-level features (HLF) listed in Table~\ref{tab:jet_features} and first used in a previous version of the algorithm, described in Ref.~\cite{Sirunyan:2017ezt}.
To isolate the effects of the different architecture, the DDB+ tagger uses the same inputs as the IN tagger, while retaining the architecture of the DDB tagger.
The charged particles (SVs) are sorted in descending order of the 2D impact parameter significance (2D flight distance significance) and only the first 60 (5) are considered.

\section{The interaction network model}
\label{sec:in}

The IN is based on two input collections comprising $N_p$ particles, each represented by a feature vector of length $P$, and $N_v$ vertices, each represented by a feature vector of length $S$.
Although kinematic features of neutral particles could also be taken into account with an additional input graph, we verified that doing so does not significantly improve the performance for this task as shown in Sec.~\ref{sec:results}.
Further, excluding neutral particles has the benefit of improved robustness to pileup.
For a single jet, the input consists of an $X$ and a $Y$ matrix, with sizes $P \times N_p$ and $S \times N_v$, respectively.
The $X$ matrix contains the input features (columns) of the charged particles (rows), while the $Y$ matrix contains the input features of the SVs.

\begin{figure}[b!]
\centering
\ifpreprint{
  \includegraphics[width=0.45\columnwidth]{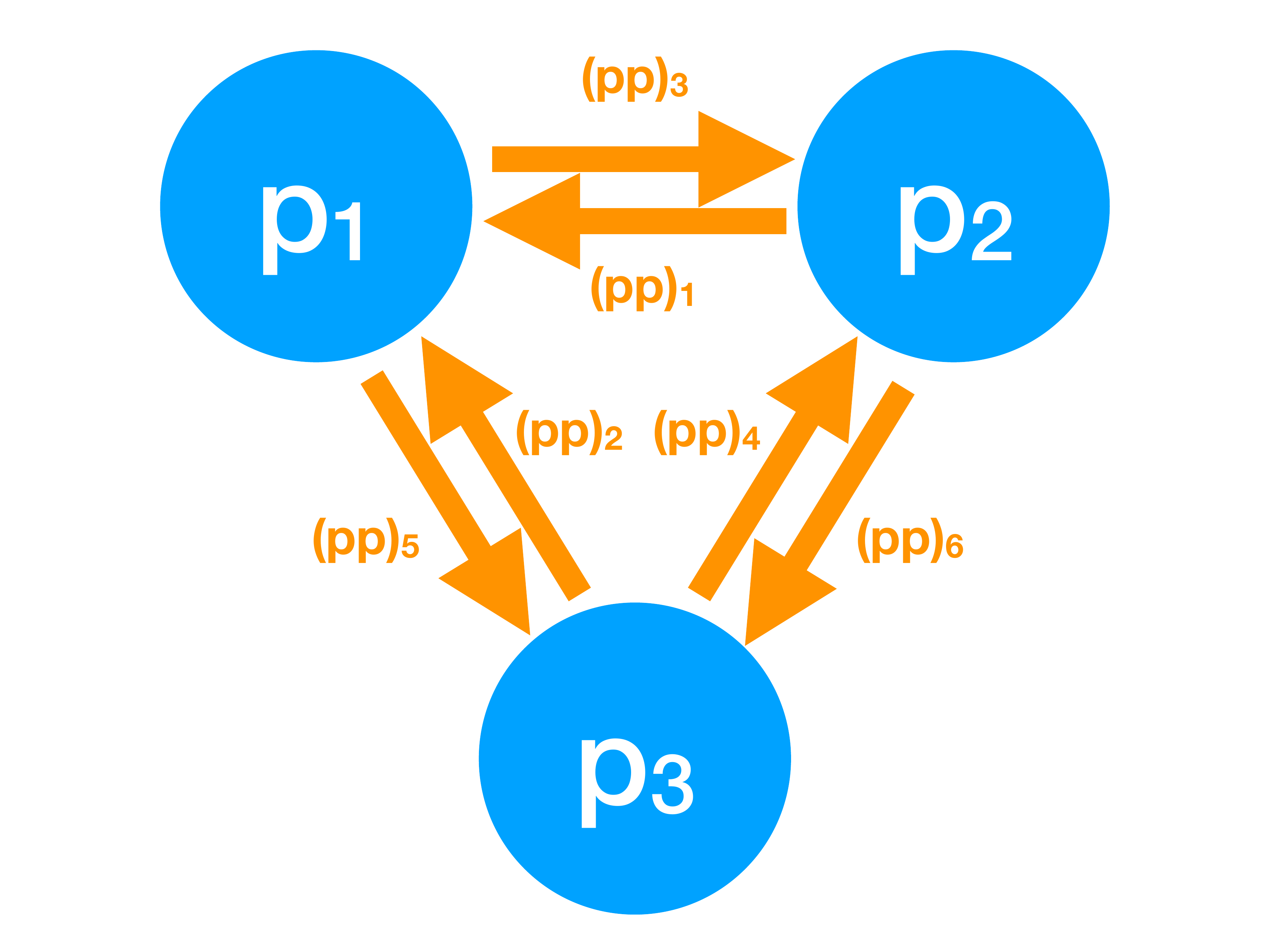}
  \includegraphics[width=0.45\columnwidth]{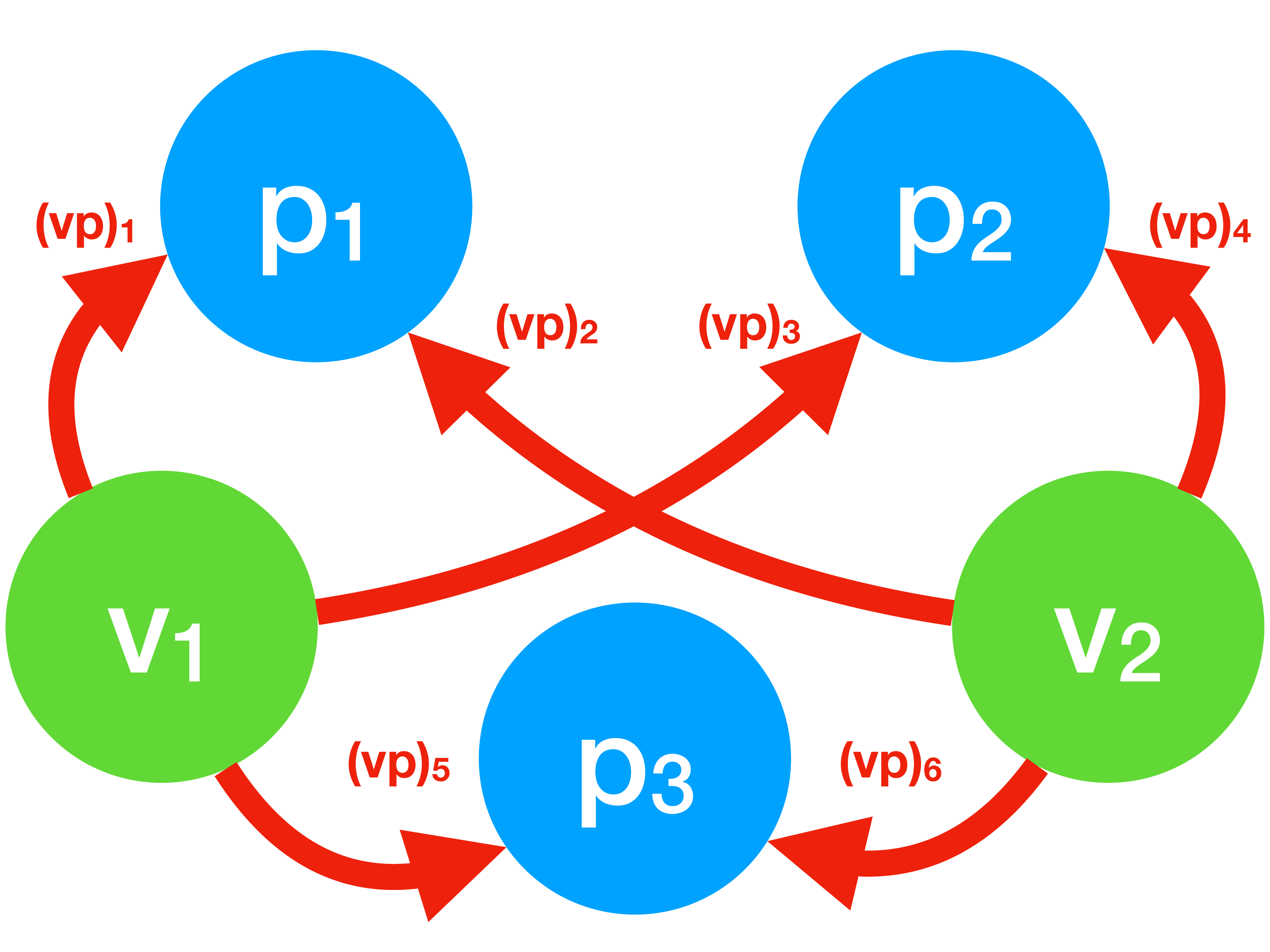}}
{  \includegraphics[width=\columnwidth]{3graph.pdf}\\
  \includegraphics[width=\columnwidth]{3graph_sv.pdf}
}
  \caption{Two example graphs with 3 particles and 2 vertices and the corresponding edges.}
  \label{fig:3graph}
\end{figure}

A particle graph $\mathcal{G}_p$ is constructed by connecting each particle to every other particle through $N_{pp} = N_p(N_p - 1)$ directed edges.
Similarly, a particle-vertex graph $\mathcal G_{pv}$ is constructed by connecting each vertex to each particle  through $N_{pv} = N_pN_v$ directed edges.
As described below, we only consider those edges that are received by particles because the final aggregation is performed over the particles.
These graphs are pictorially represented in Fig.~\ref{fig:3graph} for the case of three particles and two vertices.
As shown in the figure, the graph nodes and edges are arbitrarily enumerated.
The result of the graph processing is independent of the labeling order, as described below.

\clearpage

For the graph $\mathcal{G}_p$, a receiving matrix ($R_R$) and a sending matrix ($R_S$) are defined, both of size $N_p\times N_{pp}$.
The element $(R_R)_{ij}$ is set to 1 when the $i$th particle receives the $j$th edge and is 0 otherwise.
Similarly, the element $(R_S)_{ij}$ is set to 1 when the $i$th particle sends the $j$th edge and is 0 otherwise.
For the second graph, the corresponding adjacency matrices $R_K$ (of size $N_p\times N_{vp}$) and $R_V$ (of size $N_v\times N_{vp}$) are defined.
In the example of Fig.~\ref{fig:3graph}, the $R_R$, $R_S$, $R_K$, and $R_V$ matrices would be written as:
\ifpreprint{
\begin{align}
    R_R &=
    \bordermatrix{~ & (pp)_1 & (pp)_2 & (pp)_3 & (pp)_4 & (pp)_5 & (pp)_6\cr
                  p_1 & 1 & 1 & 0 & 0 & 0 & 0\cr
                  p_2 & 0 & 0 & 1 & 1 & 0 & 0\cr
                  p_3 & 0 & 0 & 0 & 0 & 1 & 1},\\
     R_S &=
    \bordermatrix{~ & (pp)_1 & (pp)_2 & (pp)_3 & (pp)_4 & (pp)_5 & (pp)_6\cr
                  p_1 & 0 & 0 & 1 & 0 & 1 & 0\cr
                  p_2 & 1 & 0 & 0 & 0 & 0 & 1\cr
                  p_3 & 0 & 1 & 0 & 1 & 0 & 0},\\
    R_K &=
    \bordermatrix{~ & (vp)_1 & (vp)_2 & (vp)_3 & (vp)_4 & (vp)_5 & (vp)_6\cr
                  p_1 & 1 & 1 & 0 & 0 & 0 & 0\cr
                  p_2 & 0 & 0 & 1 & 1 & 0 & 0\cr
                  p_3 & 0 & 0 & 0 & 0 & 1 & 1},\\
     R_V &=
    \bordermatrix{~ & (vp)_1 & (vp)_2 & (vp)_3 & (vp)_4 & (vp)_5 & (vp)_6\cr
                  v_1 & 1 & 0 & 1 & 0 & 1 & 0\cr
                  v_2 & 0 & 1 & 0 & 1 & 0 & 1}.
\end{align}
}{
\begin{widetext}
\begin{align}
    R_R &= 
    \bordermatrix{~ & (pp)_1 & (pp)_2 & (pp)_3 & (pp)_4 & (pp)_5 & (pp)_6\cr
                  p_1 & 1 & 1 & 0 & 0 & 0 & 0\cr
                  p_2 & 0 & 0 & 1 & 1 & 0 & 0\cr
                  p_3 & 0 & 0 & 0 & 0 & 1 & 1},\\
     R_S &=
    \bordermatrix{~ & (pp)_1 & (pp)_2 & (pp)_3 & (pp)_4 & (pp)_5 & (pp)_6\cr
                  p_1 & 0 & 0 & 1 & 0 & 1 & 0\cr
                  p_2 & 1 & 0 & 0 & 0 & 0 & 1\cr
                  p_3 & 0 & 1 & 0 & 1 & 0 & 0},\\
    R_K &= 
    \bordermatrix{~ & (vp)_1 & (vp)_2 & (vp)_3 & (vp)_4 & (vp)_5 & (vp)_6\cr
                  p_1 & 1 & 1 & 0 & 0 & 0 & 0\cr
                  p_2 & 0 & 0 & 1 & 1 & 0 & 0\cr
                  p_3 & 0 & 0 & 0 & 0 & 1 & 1},\\
     R_V &= 
    \bordermatrix{~ & (vp)_1 & (vp)_2 & (vp)_3 & (vp)_4 & (vp)_5 & (vp)_6\cr
                  v_1 & 1 & 0 & 1 & 0 & 1 & 0\cr
                  v_2 & 0 & 1 & 0 & 1 & 0 & 1}.
\end{align}
\end{widetext}
}
Each column of an adjacency matrix corresponds to a directional connection from one particle to another, $(pp)_i$, or from a vertex and to a particle, $(vp)_j$.
Column entries that are 1 in a given row in the receiving matrix $R_R$ indicate that the corresponding particle receives that connection.
Likewise, if a column entry is 1 in a given row in the sending matrix $R_S$, the corresponding particle is the sender for that connection.
Because the fully connected particle graph we consider has no self-connections, i.e. no particle sends and receives the same connection, the rows of $R_R$ and $R_S$ do not share any of the same nonzero column entries.
For the $R_R$ and $R_V$ adjacency matrices, we only consider those connections that are sent to particles because the final aggregation is performed over the particles.
We tested a version of the IN architecture in which we considered connections that are sent to vertices as well and aggregated separately before being processed by the final network, but found no significant improvement.

The data flow of the IN model is pictorially represented in Fig.~\ref{fig:flow}.
The input processing starts by creating the $2P\times N_{pp}$ particle-particle interaction matrix $B_{pp}$ and the $(P+S)\times N_{vp}$ particle-vertex interaction matrix $B_{vp}$ defined as:
\begin{align}
  B_{pp} &= \begin{pmatrix}X \cdot R_R \\
  X \cdot R_S\end{pmatrix},\\
  B_{vp} &= \begin{pmatrix}X \cdot R_K \\
  Y \cdot R_V\end{pmatrix},
\end{align}
where $\cdot$ indicates the ordinary matrix product.
Each column of $B_{pp}$ consists of the $2P$ features of the sending and receiving nodes of each particle-particle interaction, while each column of $B_{vp}$ consists of the $P+S$ features of each particle-vertex one.

Processing each column of $B_{pp}$ by the function $f_R^{pp}$, one builds an internal representation of the particle-particle interaction with a function $f_R^{pp}: \mathbb{R}^{2P} \mapsto \mathbb{R}^{D_E}$, where $D_E$ is the size of the internal representation.
This results in an \emph{effect matrix} $E_{pp}$ with dimensions $D_E\times N_{pp}$.
We similarly build the $E_{vp}$ matrix, with dimensions $D_E\times N_{vp}$, using a function $f_R^{vp}:\mathbb{R}^{P+S} \mapsto \mathbb{R}^{D_E}$.

\begin{figure*}[t!]
  \centering
  \includegraphics[width=\textwidth]{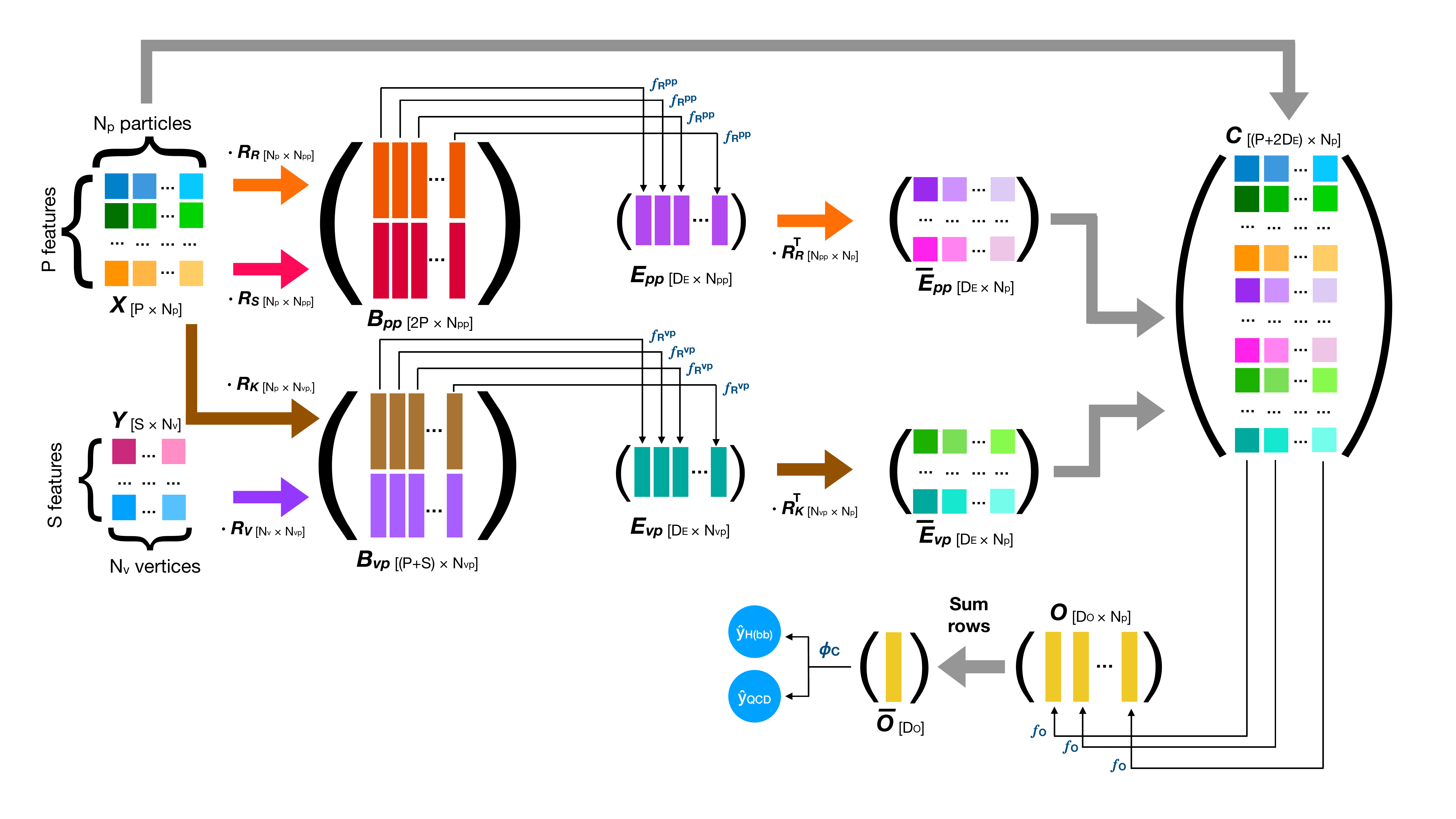}
  \caption{Illustration of the IN classifier.
    The particle feature matrix $X$ is multiplied by the receiving and sending matrices $R_R$ and $R_S$ to build the particle-particle interaction feature matrix $B_{pp}$.
    Similarly, the particle feature matrix $X$ and the vertex feature matrix $Y$ are multiplied by the adjacency matrices $R_K$ and $R_V$, respectively, to build the particle-vertex interaction feature matrix $B_{vp}$.
    These pairs are then processed by the interaction functions $f_R^{pp}$ and $f_R^{vp}$, and the post-interaction function $f_O$, which are expressed as neural networks and learned in the training process.
    This procedure creates a learned representation of each particle's post-interaction features, given by $N_p$ vectors of size $D_O$.
    The $N_p$ vectors are summed, giving $D_O$ features for the entire jet, which is given as input to a classifier $\phi_C$, also represented by a neural network.
    More details on the various steps are given in the text.\label{fig:flow}}
\end{figure*}

We then propagate the particle-particle interactions back to the particles receiving them, by building  $\overline{E}_{pp} = E_{pp} R_R^\top$ with dimension $D_E \times N_p$.
We also build $\overline{E}_{vp} = E_{vp} R_V^\top$ with dimension $D_E \times N_p$, which collects the information of the particle-vertex interactions for each particle and across all of the vertices.

The next step consists of building the $C$ matrix, with dimensions $(P + 2D_E)\times N_p$, by combining the input information for each particle ($X$) with the learned representation of the particle-particle ($\overline{E}_{pp}$) and particle-vertex ($\overline{E}_{vp}$) interactions:
\begin{align}
C &= \begin{pmatrix}X\\\overline{E}_{pp}\\\overline{E}_{vp}\end{pmatrix}.
\end{align}
The final aggregator combines the input and interaction information to build the postinteraction representation of the graph, summarized by the matrix $O$, with dimensions $D_O \times N_p$.
The aggregator consists of a function $f_O: \mathbb{R}^{P + 2D_E} \mapsto \mathbb{R}^{D_O}$, which computes the elements of the $O$ matrix
The elements of the $O$ matrix are computed by a function $f_O: \mathbb{R}^{P + 2D_E} \mapsto \mathbb{R}^{D_O}$, which returns the postinteraction representation for each of the input nodes.
As is done for $f_R^{pp}$ and $f_R^{vp}$, $f_O$ is applied to each column of $C$.

We stress the fact that the by-column processing applied by the $f_R^{pp}$,  $f_R^{vp}$, and $f_O$ functions and the sum across interactions by defining the $\overline{E}_{pp}$ and $\overline{E}_{vp}$ matrices are essential ingredients to make the outcome of the IN tagger independent of the order used to label the $N_p$ input particles and $N_v$ input vertices.
In other words, while the representations of the $R_R$, $R_S$, $R_K$, and $R_V$ matrices depend on the adopted labeling convention, the final representation of each particle does not.

The learned representation of the post-interaction graph, given by the elements of the $O$ matrix, can be used to solve the specific task at hand.
Depending on the task, the final function that computes the classifier output may be chosen to preserve the permutation invariance of the input particles and vertices.
In this case, we first sum along each row (corresponding to a sum over particles) of $O$ to produce a feature vector $\overline{O}$ with length $D_O$ for the jet as a whole.
This is passed to a function $\phi_C: \mathbb{R}^{D_O} \mapsto \mathbb{R}^{N}$, which produces the output of the classifier.

The training of the IN is performed with the CMS open simulation with 2016 conditions.
The input dataset is split into training, validation, and test samples with percentages of 80\%, 10\%, and 10\%, respectively.

We use \textsc{PyTorch}~\cite{ketkar2017introduction,NEURIPS2019_9015} to implement and train the classifier on one NVIDIA GeForce GTX 1080 GPU.
We also convert the interaction network into a \textsc{TensorFlow} model, as discussed in Appendix~\ref{sec:tensorflowconversion}.
The model is implemented with each of $f^{pp}_R$ and $f^{vp}_R$ expressed as a sequence of 3 dense layers of sizes $(60, 30, 20)$ with a rectified linear unit (ReLU) activation function after each layer.
The function $f_O$ is a similar sequence of dense layers of sizes $(60, 30, 24)$ with ReLU activations.
We use up to $N_p = 60$ charged particles and $N_v = 5$ secondary vertices as inputs to the IN tagger.
Given the size of these layers, the total number of trainable parameters is 18,144.
We train the model using the Adam optimizer~\cite{kingma2014adam} with an initial learning rate of $10^{-4}$ and a batch size of 128 for up to 200 epochs, enforcing early stopping~\cite{Yao2007} on the validation loss with a patience of 5 epochs.
The size of the batch is constrained by the required memory utilization of the GPU.
The training takes approximately 25 minutes per epoch on the GPU and stopped after 110 epochs.

For the baseline algorithm, we minimize the categorical cross-entropy loss function for this classification task $L_{\mathrm{C}}$ and let the network exploit all of the discriminating information in the dataset.

To determine the impact of neutral particles, we also train an augmented all-particle IN model, which consumes an additional input set with 10 kinematic features for up to 100 charged or neutral particles, listed in Table~\ref{tab:part_features}.
This additional input set is processed by the model in a similar way to the SV input set: the set of all particles is fully connected to the set of charged particles. 
The effect matrix for these interactions is computed by an independent neural network and then appended to an enlarged $C$ matrix, now of size $(P+3D_E)\times N_p$, before being processed by the network $f_O$.
The remaining steps of the model proceed as described above.
The total number of trainable parameters for this model is 24,254.

\section{Decorrelation with the jet mass}
\label{sec:decorrelation}
Many possible applications of a jet tagging algorithm would require the final score to be uncorrelated from the jet mass, so that a selection based on the tagger score does not change the jet mass distribution. This is particularly relevant for the background distribution, but is required to some extent also for the signal one.
Several techniques exist to deliver a tagger with minimal effects on the jet mass distribution.
For taggers based on high-level features, one could remove those features more correlated to the jet mass or divide those correlated features by the jet mass.
For taggers based on a more {\it raw} representation of the jet (as in this case), one could perform an adversarial training~\cite{2014arXiv1412.4446A,JMLR:v17:15-239,Louppe:2016ylz,Shimmin:2017mfk,ATL-PHYS-PUB-2018-014}.
One could also reweight or remove background events such that the background $\mSD$ distribution is indistinguishable from the signal $\mSD$ distribution~\cite{Bradshaw:2019ipy}.
Finally, one could also define a mass-dependent threshold based on simulation as in the ``designing decorrelated taggers'' (DDT) procedure proposed in Ref.~\cite{Dolen:2016kst}.
We test and compare all three methods in App.~\ref{sec:moredecor}.
We found the DDT method to be the most robust and performant deocorrelation procedure.
As such, we use it as the nominal decorrelation method in the following results.

\subsection{Designing decorrelated taggers}
\label{sec:ddt}

Following the DDT procedure~\cite{Dolen:2016kst}, the tagger threshold for a given false positive rate (FPR) or ``working point'' is determined as a function of $\mSD$.
By creating a $\mSD$-dependent tagger threshold, the background jet $\mSD$ distribution for events passing and failing this threshold can be made identical.
In practice, this is done by considering the distribution of the network score versus the jet $\mSD$ for the training dataset.
A quantile regression was used to find the threshold on the network score as a function of $\mSD$ distribution that would correspond to a fixed quantile (the chosen $1-$FPR value).
By construction, this procedure results in near-perfect mass decorrelation.

In this case, a gradient boosted regressor~\cite{friedman2002stochastic, friedman2001greedy} with the following parameters was used:
\begin{itemize}
    \item $\alpha$-quantile of $1 - \mathrm{FPR}$,
    \item number of estimators of 500,
    \item minimum number of samples at a leaf node of 50,
    \item minimum number of samples to split an internal node of 2500,
    \item maximum depth of 5,
    \item validation set of 20\%,
    \item early stopping with tolerance of 10.
\end{itemize}

\section{Deep double-b tagger models}
\label{sec:ddb}

The DDB tagger is a convolutional and recurrent neural network model developed by CMS~\cite{CMS-Hbb-baseline} to identify boosted $\PH \to \bbbar$ jets.
We reconstruct this model based on publicly available information from the CMS Collaboration as follows.
The model takes as input 27 HLFs used in Ref.~\cite{Sirunyan:2017ezt}, as well as 8 particle-specific features of up to 60 charged particles, and 2 properties of up to 5 SVs associated with the jet (see Appendix~\ref{sec:features}).
Each block of inputs is treated as a one-dimensional list, with batch normalization~\cite{batchnorm} applied directly to the input layers.
For each collection of charged particles and SVs, separate 1D convolutional layers~\cite{conv1d}, with a kernel size of 1, are applied: 2 hidden layers with 32 filters each and ReLU~\cite{agarap2018learning} activation.
The outputs are then separately fed into two gated recurrent units (GRUs) with 50 output nodes each and ReLU activations.
Finally, the GRU outputs are concatenated with the HLFs and processed by a dense layer with 100 nodes and ReLU activation, and another final dense layer with 2 output nodes with softmax activation.
Dropout~\cite{dropout} (with a rate of 10\%) is used in each layer to prevent overfitting.
The nominal DDB tagger model has 40,344 trainable parameters, 32\% of which are found in the fully connected layers.

We define a variant of this model, the DDB+ model, which takes as input all 30 features of charged particles and all 14 features of the SVs.
In this variant, we do not consider the HLFs.
Thus, the final dense layer only receives the GRU outputs from processing the low-level charged particle and SV information.
This extended DDB+ tagger algorithm has 38,746 trainable parameters.
The number of parameters is less overall because the increase in the size of the convolutional and recurrent layers is compensated by the decrease in the size of the fully connected layers.

\begin{figure}[b!]
  \centering
   \includegraphics[width=\columnwidth]{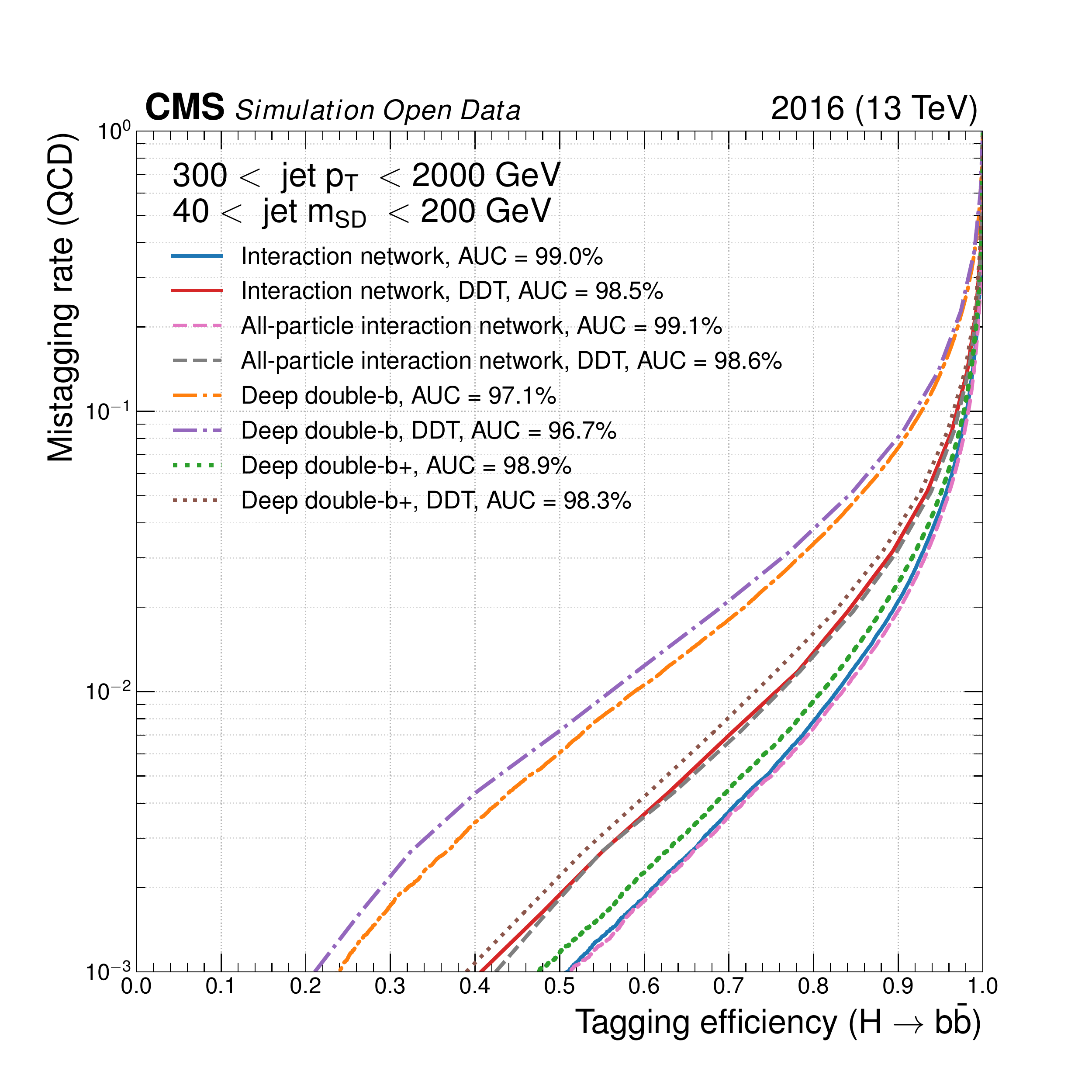}
    \caption{Performance of the IN, all-particle IN, DDB, and DDB+ algorithms quantified with a ROC curve of FPR (QCD mistagging rate) versus TPR ($\PH\to\bbbar$ tagging efficiency).
    The performance of each baseline algorithm is compared to that of the algorithms after applying the DDT procedure to decorrelate the tagger score from the jet mass.
      This decorrelation results in a smaller TPR for a given FPR.}
    \label{fig:in_performance}
\end{figure}

\begin{figure}[b!]
  \centering
  \ifpreprint{
   \includegraphics[width=0.45\columnwidth]{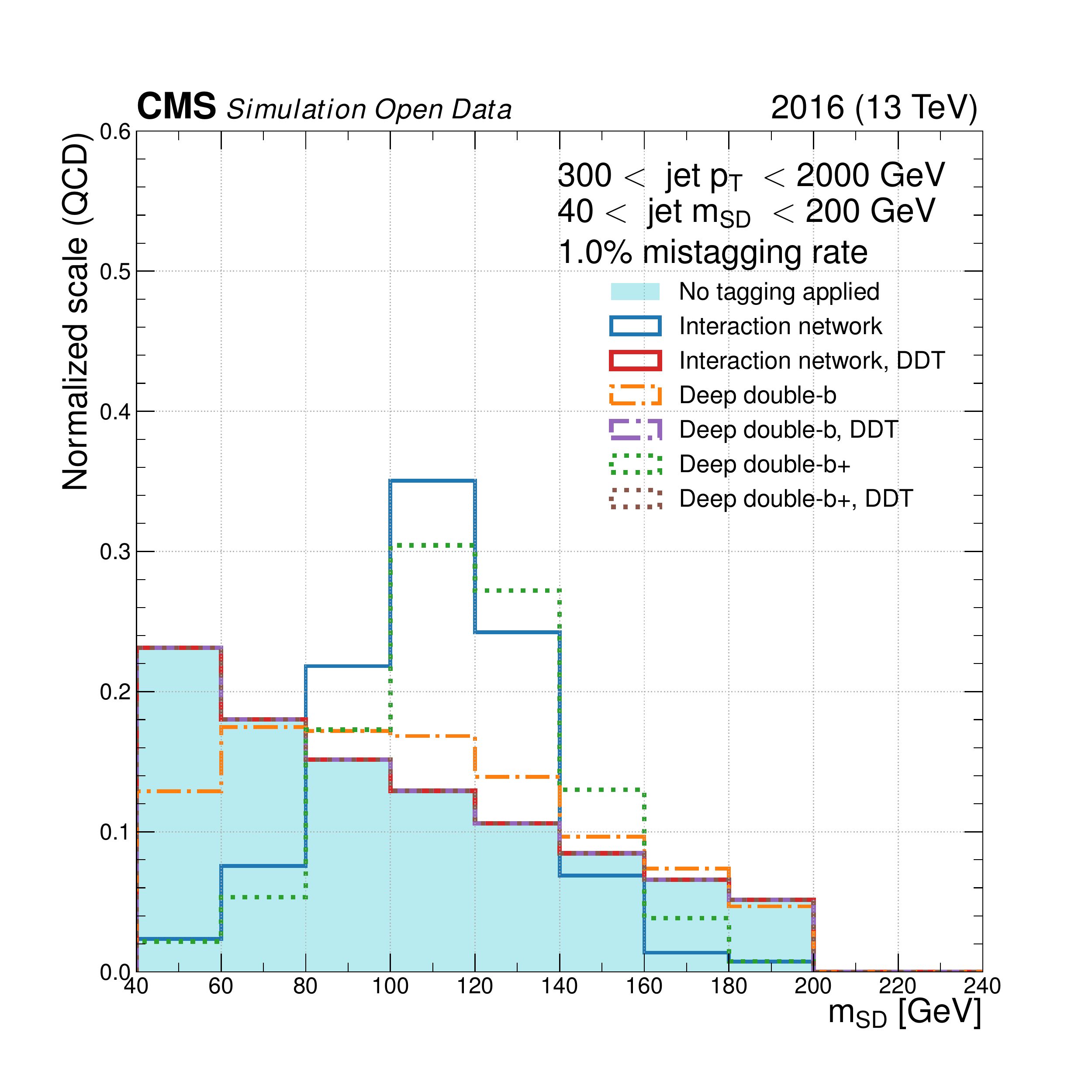}
   \includegraphics[width=0.45\columnwidth]{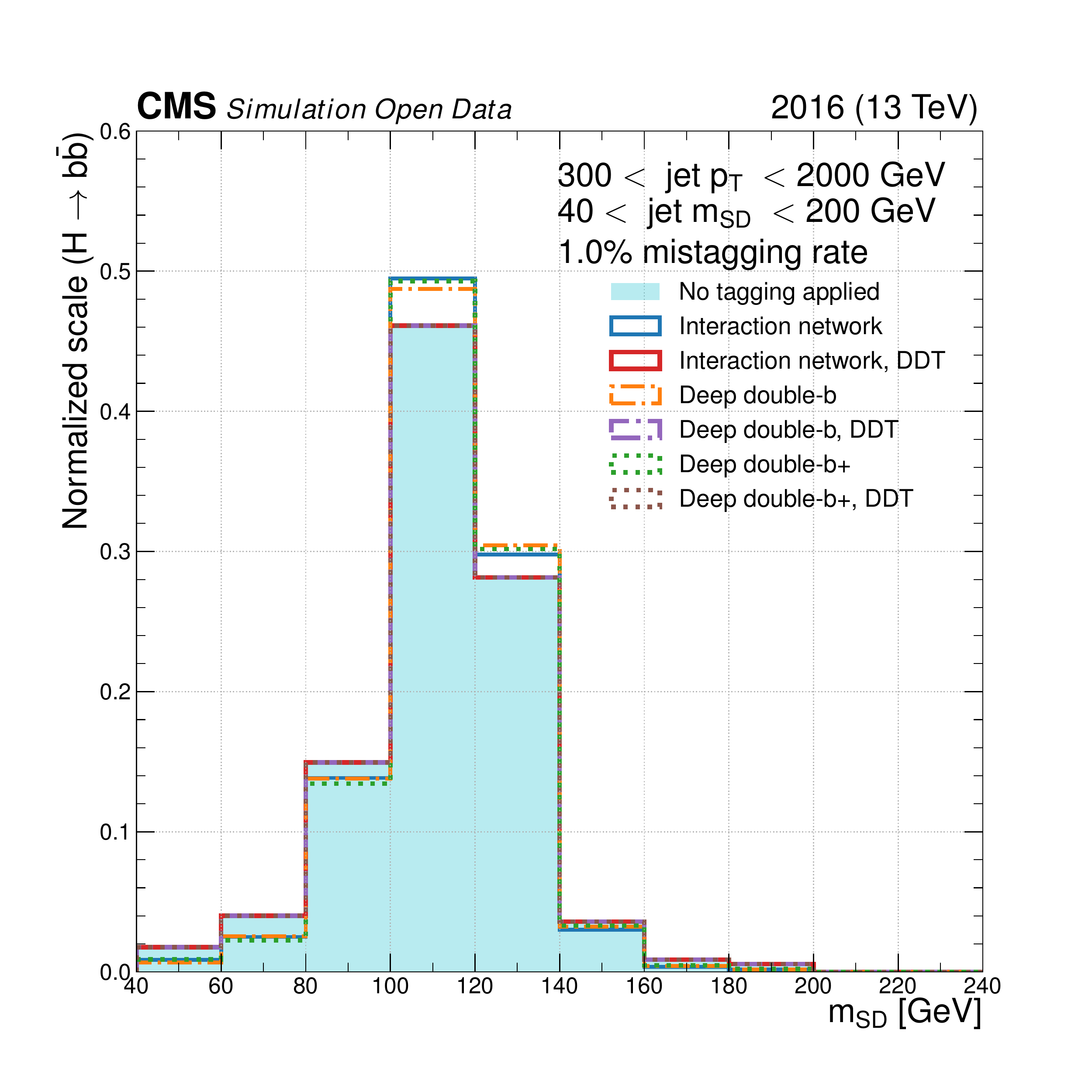}
}{
   \includegraphics[width=\columnwidth]{M_sculpting_tagHbb_QCD_final.pdf}\\
   \includegraphics[width=\columnwidth]{M_sculpting_tagHbb_Hbb_final.pdf}
}
   \caption{ An illustration of the ``sculpting'' of the background jet mass distribution (\ifpreprint{left}{top}) and the signal jet mass distribution (\ifpreprint{right}{bottom}) after applying a threshold on the tagger score corresponding to a 1\% FPR for several different algorithms.
     The unmodified interaction network is highly correlated with the jet mass, but after applying the methods described in the text, the correlation is reduced for the background while the peak of the signal distribution is still retained.  }
    \label{fig:mass_sculpting}
\end{figure}

We train the DDB and DDB+ models using the CMS open simulation dataset with \textsc{Keras}~\cite{chollet2015keras} over up to 200 epochs with an early stopping patience of 5 epochs and a batch size of 4096 using the Adam optimizer with an initial learning rate of $10^{-3}$.
For both models, one training epoch takes about 3 minutes and training stops after approximately 50 epochs.
In this case, the larger batch size is possible due to the smaller GPU memory utilization of the model during training.
We find consistent performance for different batch size choices with no evidence of overfitting with larger batch sizes.

In order to decorrelate the tagger output from the jet mass, we use the same DDT procedure described in Sec.~\ref{sec:ddt} applied to both the DDB and DDB+ taggers.

\begin{table*}[t!]
\centering
\caption{\label{tab:perf} Performance metrics of the different baseline and decorrelated models, including accuracy, area under the ROC curve, background rejection at a true positive rate of 30\% an 50\%, and true positive rate and mass decorrelation metric $1/D_\mathrm{JS}$ at a false positive rate of 1\%. For the DDT models, the corresponding accuracy is listed for the tagger after the decorrelation is performed for a FPR of 50\%.}
\resizebox{\textwidth}{!}{
\begin{ruledtabular}
\begin{tabular}{lccccccc}
\multirow{2}{*}{Model} & \multirow{2}{*}{Parameters} &  \multirow{2}{*}{Accuracy} & \multirow{2}{*}{AUC} & $1/\varepsilon_\mathrm{b}$& $1/\varepsilon_\mathrm{b}$& $\varepsilon_\mathrm{s}$ &  $1/D_\mathrm{JS}$\\
 & & & & @ $\varepsilon_\mathrm{s}=30\%$ &  @ $\varepsilon_\mathrm{s}=50\%$ &@ $\varepsilon_\mathrm{b}=1\%$ & @ $\varepsilon_\mathrm{b}=1\%$ \\\hline
\multicolumn{8}{c}{Baseline models}  \\\hline
Interaction network & \textbf{18,144} &  \textbf{95.5\%}  & \textbf{99.0\%} & \textbf{4616.9} & \textbf{1028.8} & \textbf{82.8\%} & 4.5 \\ 
Deep double-$\PQb$ & 40,344 & 91.7\% & 97.2\% & 578.0 & 165.3 & 60.6\% & \textbf{75.3} \\ 
Deep double-$\PQb$+ & 38,746 & 95.3\% & 98.8\% & 3863.1 & 852.7 & 81.5\% & 4.4 \\\hline  
\multicolumn{8}{c}{Decorrelated models}  \\\hline
Interaction network, DDT & \textbf{18,144} & \textbf{93.2\%} & \textbf{98.5\%} & \textbf{2258.7} & \textbf{540.0} & \textbf{75.6\%} & \text{29,265.3}\\
Deep double-$\PQb$, DDT & 40,344 & 86.8\% & 96.7\% & 456.6 & 136.8 & 55.9\% & \textbf{48,099.0}\\
Deep double-$\PQb$+, DDT & 38,746 & 92.9\% & 98.3\% & 1973.8 & 466.6 & 72.9\% & 15,171.2 \\
\end{tabular}
\end{ruledtabular}
}
\end{table*}

\section{Results}
\label{sec:results}

In Fig.~\ref{fig:in_performance} the performance of the IN, all-particle IN, DDB, and DDB+ algorithms are quantified in a ROC curve.
The axes are the TPR, or $\PH\to\bbbar$ tagging efficiency and the false positive rate, or QCD mistagging rate.
As shown in Fig.~\ref{fig:in_performance}, the IN provides an improved performance with respect to the DDB and DDB+ taggers.
At a 1\% FPR, the IN tagger outperforms the DDB and DDB+ taggers by 37\% and 2\% in TPR, respectively.
 Likewise, at a 50\% TPR, the IN tagger yields a factor of 6 or 1.2 better background rejection (1/FPR) than the DDB or DDB+ tagger, respectively.
Thus, while the additional inputs provide a significant improvement for the DDB+ model, the IN architecture is also important to achieve a better performance with significantly less parameters than the DDB+ model.

We verified that one could match the performance obtained by the IN with a DDB-inspired architecture and expanding the model size.
With 150,786 trainable parameters, a DDB architecture achieves the same performance as the IN at the cost of 8 times more parameters.
Because of this the IN model holds an advantage in terms of memory usage during inference over this alternative model.

Figure~\ref{fig:in_performance} also shows that there is only a modest improvement in the AUC and accuracy by including information in the IN model from neutral particles. 
For this reason and to preserve robustness to increased pileup, in the following results, we consider the original IN model that excludes neutral particles.

Figure~\ref{fig:mass_sculpting} shows an illustration of how the signal and background jet mass distributions change after applying a threshold on the different baseline and DDT-decorrelated tagger scores.
Following Ref.~\cite{ATL-PHYS-PUB-2018-014}, we quantify the impact of these algorithms on the mass decorrelation by computing the Jensen-Shannon (JS) divergence:
\begin{align}
    D_{\mathrm{JS}}(P \parallel Q) &= \frac{1}{2}D_{\mathrm{KL}}(P \parallel M) + \frac{1}{2}D_{\mathrm{KL}}(Q\parallel M),
\end{align}
where $M= \frac{1}{2}(P+Q)$ is the average of the normalized $\mSD$ distributions of the background jets passing ($P$) and failing ($Q$) a given tagger score and $D_{\mathrm{KL}}(P \parallel Q  ) = \sum_{i} P_i \log (P_i/Q_i)$ is the Kullback-Leibler (KL) divergence.
Larger values of the metric $1/D_\mathrm{JS}$ correspond to a better decorrelation.

\begin{figure}[b!]
  \centering
  \includegraphics[width=\columnwidth]{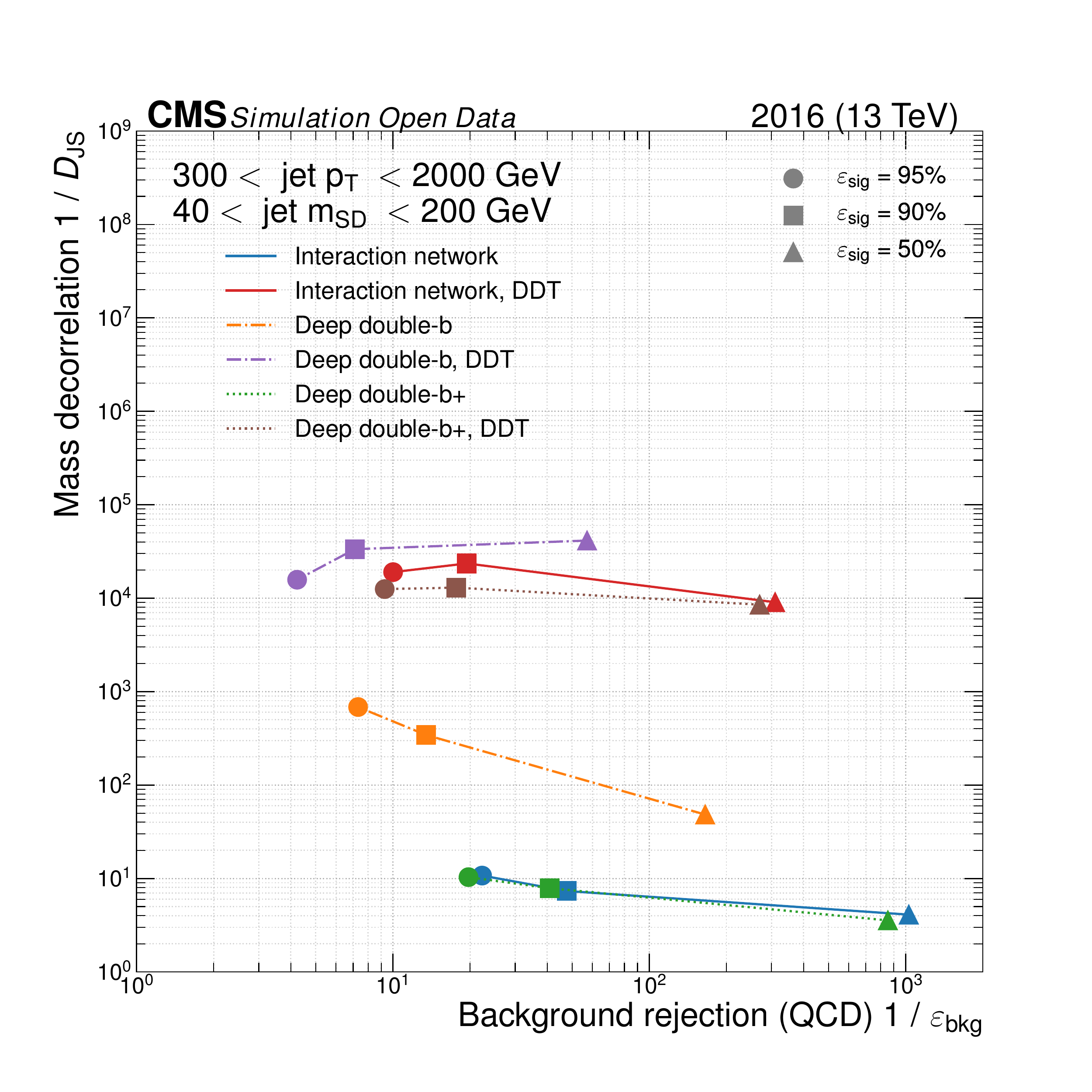}
    \caption{The mass decorrelation metric $1/D_\mathrm{JS}$ as a function of background rejection for the baseline and decorrelated IN, DDB, and DDB+ taggers.
      The decorrelation is quantified as the inverse of the JS divergence between the background mass distribution passing and failing a given threshold cut on the classifier score.
      Greater values of this metric correspond to better mass decorrelation.
      The background rejection is quantified as the inverse of the FPR, while the signal efficiency is equal to the TPR.\label{fig:JSD}}
\end{figure}

After applying the mass decorrelation techniques, the performance of each of the taggers worsens slightly but the IN algorithm still significantly outperforms the DDB and DDB+ taggers.
Figure~\ref{fig:JSD} displays the trade-off between the background rejection and $1/D_\mathrm{JS}$ at different TPRs for the baseline and DDT-decorrelated algorithms.
At a 50\% TPR, the decorrelated IN algorithm achieves a significantly better $1/D_\mathrm{JS}$ by a factor of about 2,200 while the background rejection decreases by a factor of about 3.3 compared to the baseline IN algorithm.
At a 1\% FPR, the DDT-decorrelated IN tagger has a TPR of 75.6\% compared to the DDT-decorrelated DDB (DDB+) tagger with a 55.9\% (72.9\%) TPR, corresponding to an improvement of 35\% (4\%).
Table~\ref{tab:perf} summarizes different performance metrics for the three considered models and their decorrelated versions.
For the DDT models, the corresponding accuracy is listed for the tagger after the decorrelation is performed for a FPR of 50\%.

To quantify the dependence on the number of pileup interactions, Fig.~\ref{fig:in_pileup} shows the performance of the different algorithms as a function of the number of primary vertices in the event, which scales linearly with the number of pileup collisions.
Using only charged particles and secondary vertices as input, the IN tagger is robust against an increasing number of pileup interactions, exhibiting behavior similar to the DDB and DDB+ taggers.

\begin{figure}[t!]
  \centering
    \includegraphics[width=\columnwidth]{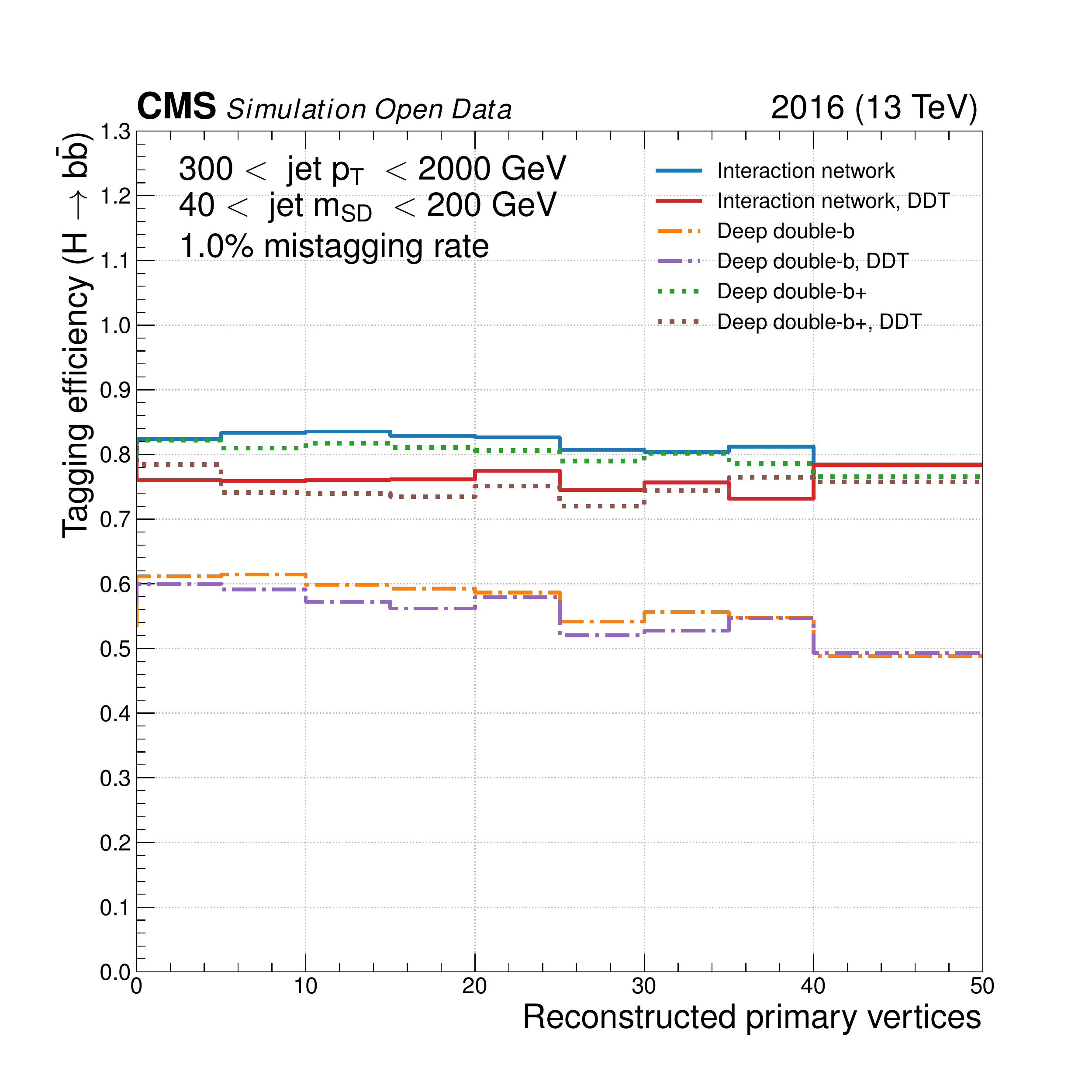}
    \caption{TPR of the baseline and decorrelated IN, DDB, and DDB+ taggers as a function of the number of reconstructed PVs for a 1\% FPR.\label{fig:in_pileup}}
\end{figure}

\section{Conclusions}
\label{sec:conclusions}

We presented a novel technique using a graph representation of the jet's constituents and secondary vertices based on an interaction network to identify Higgs bosons decaying to bottom quark-antiquark pairs ($\PH\to\bbbar$) in LHC collisions.
This model can operate on a variable number of jet constituents and secondary vertices and does not depend on the ordering schemes of these objects.
The interaction network was trained on an open simulation dataset released by the CMS Collaboration in the CERN Open Data Portal.
A significant improvement in performance is observed with respect to two alternative taggers based on the deep double-$\PQb$ tagger created by the CMS Collaboration.
By design, the interaction network uses extended low-level input features for particles and vertices, offers a more flexible representation of jet data, and is robust against the noise generated by pileup collisions.
Even when trained with the same set of input features, the interaction network architecture outperforms the deep double-$\PQb$ architecture.
Thus, while part of the improvement is due to the extended input representation, additional improvement comes from the interaction network architecture, despite using on half as many parameters.
The interaction network algorithm implementation and its training code are available at Ref.~\cite{INcode}.

Together with the best-performing models, we presented additional models, obtained by applying different decorrelation techniques between the network score and the jet-mass distribution.
This was done to minimize the selection bias of the classifier output towards any values of the jet mass, which would make the algorithms suitable for physics analyses relying on the jet mass as a discriminating variable.
As expected, the decorrelation procedure results in a reduction of the $\PH\to\bbbar$ identification performance.
Nevertheless, the decorrelated interaction network model outperforms the decorrelated deep double-$\PQb$ models.

Once applied to a full data analysis, this graph-based tagging algorithm could contribute a substantial improvement to the experimental precision of $\PH\to\bbbar$ measurements, including those sensitive to beyond the standard model physics and the Higgs boson self-coupling.
These results motivate further exploration of applications based on interaction networks (and graph neural networks in general) for object tagging and other similar tasks in experimental high energy physics.

\begin{acknowledgments} 

This work was possible thanks to the commitment of the CMS Collaboration to release its simulation data through the CERN Open Data Portal.
We would like to thank our CMS colleagues and the CERN Open Data team for their effort to promote open access to science.
In particular, we thank Kati Lassila-Perini for her precious help.

We are grateful to Caltech and the Kavli Foundation for their support of undergraduate student research in cross-cutting areas of machine learning and domain sciences.
This work was conducted at ``\textit{iBanks},'' the AI GPU cluster at Caltech.
We acknowledge NVIDIA, SuperMicro and the Kavli Foundation for their support of \textit{iBanks}.
This project is partially supported by the United States Department of Energy, Office of High Energy Physics Research under Caltech Contract No. DE-SC0011925.
M.~P. is supported by the European Research Council (ERC) under the European Union's Horizon 2020 research and innovation program (Grant Agreement No. 772369).
J.~M.~D. is supported by Fermi Research Alliance, LLC under Contract No. DE-AC02-07CH11359 with the U.S. Department of Energy, Office of Science, Office of High Energy Physics.
E.~A.~M. is supported by the Taylor W. Lawrence Research Fellowship and Mellon Mays Research Fellowship.

\end{acknowledgments} 

\appendix
\section*{Appendix A: Additional mass decorrelation methods}
\label{sec:moredecor}

In this appendix, we describe and compare two additional mass decorrelation methods to the DDT procedure described in Section~\ref{sec:decorrelation}.
In one method, we train two neural networks simultaneously: the original classifier and an additional network intended to regress the jet mass, known as the adversary.
The original classifer is trained to maximally confuse the adversary.
After training, the effect is the classifier is not able to discriminate the jet mass.
In the other method, we train the classifier with sample weights, such that the QCD background jet mass distribution is reweighted to be identical to that of the $\PH\to\bbbar$ signal.
We then compare these procedures to the DDT method.

\subsection{Adversarial training}
The secondary adversary network is constructed that consists of three hidden layers each with 64 nodes.
The adversary is trained simultaneously with the classifier (interaction network) using the summed post-interaction feature vector $\mathbf{\overline{O}}$ as its input.
From this input, the adversary is trained to predict a one-hot encoding of the pivot feature $\mSD$, which we aim to decorrelate from the classifier output.
The chosen one-hot encoding corresponds to 40 $\mSD$ bins from 40 to 200\GeV.
The training begins by initializing the weights from the best classifier training.
The adversary is then pre-trained for 10 epochs using the Adam algorithm with an initial learning rate of $10^{-4}$.
During each epoch, the classifier is first trained by minimizing the total loss
\begin{align}
    L &= L_{\mathrm{C}} - \lambda L_{\mathrm{adversary}}.
\end{align}
Subsequently, the adversary is trained by minimizing $L_{\mathrm{adversary}}$ using only the background QCD samples.
To balance tagging performance and $\mSD$ correlation, $\lambda=10$ was chosen.

\subsection{Sample reweighting}
While adversarial training requires a complicated tuning process, sample reweighting is a simpler way to achieve the same goal.
Individual QCD events are weighted in the loss function based on their mass bin as to match the signal jet mass distribution of the training sample.
Given a background event in certain mass bin, with the number of background and signal events in that bin denoted as $N^\mathrm{bin}_\mathrm{b}$ and $N^\mathrm{bin}_\mathrm{s}$, respectively, the event is weighted by $w_\mathrm{bin}=N^\mathrm{bin}_\mathrm{s}/N^\mathrm{bin}_\mathrm{b}$.

\begin{figure}[t!]
  \centering
   \includegraphics[width=\columnwidth]{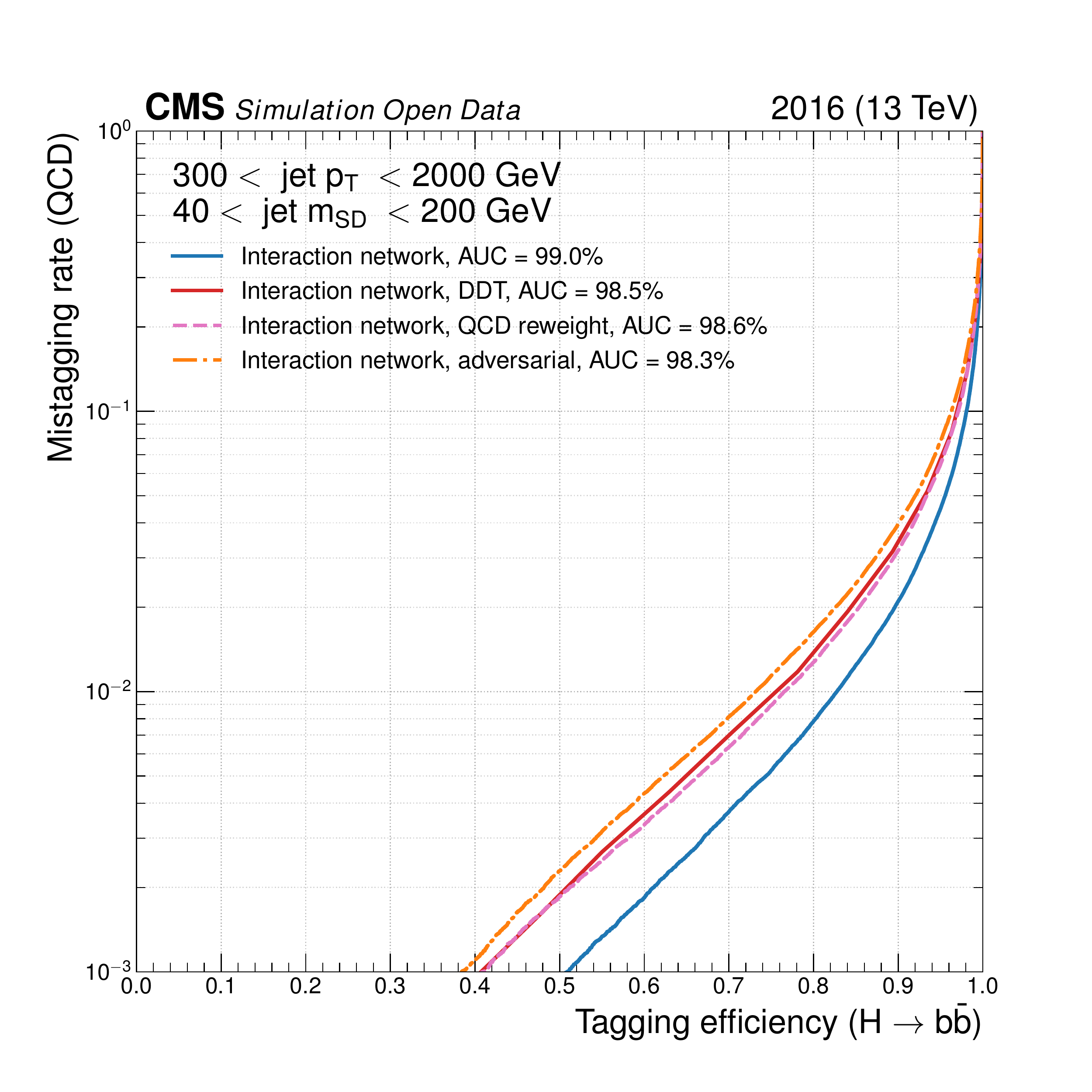}
    \caption{Performance of IN algorithm compared to the same after applying the three techniques described in the text to decrease the degree to which the tagger score is dependent on the mass of the jet.
      This results in a lower performance because the algorithm is forced to have reduced correlation with the jet mass.}
    \label{fig:in_decor}
\end{figure}

\begin{table*}[hbtp!]
\centering
\caption{\label{tab:deco} Performance metrics of the three decorrelated IN models, including accuracy, area under the ROC curve, background rejection at a true positive rate of 30\% and 50\%, and true positive rate and mass decorrelation metric $1/D_\mathrm{JS}$ at a false positive rate of 1\%.}
\resizebox{\textwidth}{!}{
\begin{ruledtabular}
\begin{tabular}{lccccccc}
\multirow{2}{*}{Model} & \multirow{2}{*}{Parameters} &  \multirow{2}{*}{Accuracy} & \multirow{2}{*}{AUC} & $1/\varepsilon_\mathrm{b}$& $1/\varepsilon_\mathrm{b}$& $\varepsilon_\mathrm{s}$ &  $1/D_\mathrm{JS}$\\
 & & & & @ $\varepsilon_\mathrm{s}=30\%$ & @ $\varepsilon_\mathrm{s}=50\%$ & @ $\varepsilon_\mathrm{b}=1\%$ & @ $\varepsilon_\mathrm{b}=1\%$ \\
\hline
Interaction network, adversarial & 18,144 & \textbf{94.6\%} & \textbf{98.6\%} & \textbf{2381.0} & \textbf{540.1} &   \textbf{76.5\%} & 124.6\\
Interaction network, QCD reweight. & 18,144 & 93.4\% & 98.3\% & 1864.9 & 436.2 &  73.2\% & 2051.0\\
Interaction network, DDT & 18,144 & 93.2\% & 98.5\% & 2258.7 & 540.0 & 75.6\% & \textbf{29,265.3}\\
\end{tabular}
\end{ruledtabular}
}
\end{table*}

\subsection{Results}

Figure~\ref{fig:in_decor} shows a comparison of the ROC curves for the baseline IN algorithm and the versions that were decorrelated using the DDT procedure, adversarial training, and sample reweighting.
Table~\ref{tab:deco} summarizes a variety of performance metrics for the decorrelated algorithms including $1/D_\mathrm{JS}$, which quantifies the success of the decorrelation procedure for a given FPR.
As shown in Fig.~\ref{fig:in_decor} and Table~\ref{tab:deco}, the DDT procedure provides the best decorrelation in terms of $1/D_\mathrm{JS}$, and comparable to the best accuracy, AUC, background rejection, and tagging efficiency.

\section*{Appendix B: Model implemented in \textsc{ONNX} and \textsc{TensorFlow}}
\label{sec:tensorflowconversion}

In order to integrate the IN algorithm into experimental workflows, it is often necessary to provide the algorithm in other formats.
For example, the CMS experimental software framework \textsc{CMSSW}~\cite{CMSTDR} currently only supports \textsc{ONNX}~\cite{onnx_2017}, \textsc{TensorFlow}~\cite{TensorFlow}, and \textsc{MXNet}~\cite{mxnet} models.
To perform this conversion, we first translate the \textsc{PyTorch} model into an \textsc{ONNX} representation using the built-in exporter.
Then the conversion to \textsc{TensorFlow} is performed with the dedicated \textsc{TensorFlow} backend for \textsc{ONNX}~\cite{onnxtf}.
The trained model in all three formats is available at Ref.~\cite{INcode}.

\section*{Appendix C: Dataset features}
\label{sec:features}

The charged particle features used by the IN and DDB taggers are listed in Table~\ref{tab:track_features}.
The SV features used by both taggers are listed in Table~\ref{tab:sv_features}, and the high-level features used only by the reconstructed DDB tagger are shown in Table~\ref{tab:jet_features}.
Finally, additional features of charged or neutral particles are listed in Table~\ref{tab:part_features} to demonstrate the change in the performance of the IN model by including neutral particles.

\newlength\trackTableWidth
\ifpreprint{\setlength\trackTableWidth{0.95\textwidth}}{\setlength\trackTableWidth{\textwidth}}

\begin{table*}[htpb!]
    \centering
    \caption{Charged particle features. The IN and DDB+ models use all of the features, while DDB algorithm uses the subset of features indicated in bold.}
    \resizebox{\trackTableWidth}{!}{
\begin{ruledtabular}
    \begin{tabular}{ll}
    \footnotesize
      Variable & Description\\\hline
      \texttt{track\_ptrel} & $\pt$ of the charged particle divided by the $\pt$ of the AK8 jet \\
      \texttt{track\_erel} & Energy of the charged particle divided by the energy of the AK8 jet \\
      \texttt{track\_phirel} & $\Delta\phi$ between the charged particle and the AK8 jet axis \\
      \texttt{track\_etarel} & $\Delta\eta$ between the charged particle and the AK8 jet axis \\
      \texttt{track\_deltaR} & $\Delta R$ between the charged particle and the AK8 jet axis \\
      \texttt{track\_drminsv} & $\Delta R$ between the associated SVs and the charged particle \\
      \texttt{track\_drsubjet1} & $\Delta R$ between the charged particle and the first soft drop subjet \\
      \texttt{track\_drsubjet2} & $\Delta R$ between the charged particle and the second soft drop subjet \\
      \texttt{track\_dz} & Longitudinal impact parameter of the track, defined as the distance of closest approach of \\
               & \quad the track trajectory to the PV projected on to the $z$ direction\\
      \texttt{track\_dzsig} & Longitudinal impact parameter significance of the track \\
      \texttt{track\_dxy} & Transverse (2D) impact parameter of the track, defined as the distance of closest approach \\
               & \quad of the track trajectory to the beam line in the transverse plane to the beam\\
      \texttt{track\_dxysig} & Transverse (2D) impact parameter  of the track \\
      \texttt{track\_normchi2} & Normalized $\chi^2$ of the track fit \\
      \texttt{track\_quality} & Track quality: \texttt{undefQuality=-1, loose=0, tight=1, highPurity=2, confirmed=3,}\\
                            & \quad \texttt{looseSetWithPV=5, highPuritySetWithPV=6, discarded=7, qualitySize=8} \\
      \texttt{track\_dptdpt} & Track covariance matrix entry ($\pt$, $\pt$) \\
      \texttt{track\_detadeta} & Track covariance matrix entry ($\eta$, $\eta$) \\
      \texttt{track\_dphidphi} & Track covariance matrix entry ($\phi$, $\phi$) \\
      \texttt{track\_dxydxy} & Track covariance matrix entry ($d_{xy}$, $d_{xy}$) \\
      \texttt{track\_dzdz} & Track covariance matrix entry ($d_{z}$, $d_{z}$) \\
      \texttt{track\_dxydz} & Track covariance matrix entry ($d_{xy}$, $d_{z}$) \\
      \texttt{track\_dphidz} & Track covariance matrix entry ($d_{\phi}$, $d_{z}$) \\
      \texttt{track\_dlambdadz} & Track covariance matrix entry ($\lambda$, $d_{z}$) \\
      {\ttfamily\bfseries trackBTag\_EtaRel} & $\Delta\eta$ between the track and the AK8 jet axis \\
      {\ttfamily\bfseries trackBTag\_PtRatio}  & Component of track momentum perpendicular to the AK8 jet axis, normalized to the track  \\
               & \quad momentum \\
      {\ttfamily\bfseries trackBTag\_PParRatio}  & Component of track momentum parallel to the AK8 jet axis, \\
               & normalized to the track momentum \\
      {\ttfamily\bfseries trackBTag\_Sip2dVal}  & Transverse (2D) signed impact parameter of the track \\
      {\ttfamily\bfseries trackBTag\_Sip2dSig} & Transverse (2D) signed impact parameter significance of the track \\
      {\ttfamily\bfseries trackBTag\_Sip3dVal} & 3D signed impact parameter of the track \\
      {\ttfamily\bfseries trackBTag\_Sip3dSig}  & 3D signed impact parameter significance of the track \\
      {\ttfamily\bfseries trackBTag\_JetDistVal}  & Minimum track approach distance to the AK8 jet axis
    \end{tabular} 
\end{ruledtabular}}
    \label{tab:track_features}
\end{table*}

\begin{table*}[htpb!]
\begin{ruledtabular}
   \begin{tabular}{ll}
      Variable & Description\\\hline
      \texttt{sv\_ptrel} & $\pt$ of the SV divided by the $\pt$ of the AK8 jet\\
      \texttt{sv\_erel} & Energy of the SV divided by the energy of the AK8 jet \\
      \texttt{sv\_phirel} & $\Delta\phi$ between the SV and the AK8 jet axis \\
      \texttt{sv\_etarel} & $\Delta\eta$ between the SV and the AK8 jet axis \\
      \texttt{sv\_deltaR} & $\Delta R$ between the SV and the AK8 jet axis \\
      \texttt{sv\_pt} & $\pt$ of the SV \\
      \texttt{sv\_mass} & Mass of the SV \\
      \texttt{sv\_ntracks} & Number of tracks associated with the SV \\
      \texttt{sv\_normchi2} & Normalized $\chi^2$ of the SV fit \\
      \texttt{sv\_costhetasvpv} & $\cos\theta$ between the SV and the PV\\
      \texttt{sv\_dxy} & Transverse (2D) flight distance of the SV \\
      \texttt{sv\_dxysig} & Transverse (2D) flight distance significance of the SV \\
      {\ttfamily\bfseries sv\_d3d} & 3D flight distance of the SV \\
      {\ttfamily\bfseries sv\_d3dsig}  & 3D flight distance significance of the SV
    \end{tabular}
\end{ruledtabular}
    \vspace{2mm}
    \caption{Secondary vertex features. The IN and DDB+ models use all of the features, while the DDB algorithm uses the subset of features indicated in bold.}
    \label{tab:sv_features}
\end{table*}

\newlength\tableWidth
\ifpreprint{\setlength\tableWidth{0.65\textwidth}}{\setlength\tableWidth{\textwidth}}

\begin{table*}[htpb!]
  \centering
    \caption{High-level features used by the DDB algorithm.}
    \resizebox{\tableWidth}{!}{
\begin{ruledtabular}
  \begin{tabular}{ll}
    Variable & Description\\\hline
    \texttt{fj\_jetNTracks} & Number of tracks associated with the AK8 jet \\
    \texttt{fj\_nSV} & Number of SVs associated with the AK8 jet ($\Delta R < 0.7$) \\
    \texttt{fj\_tau0\_trackEtaRel\_0} & Smallest track $\Delta\eta$ relative to the jet axis, associated to the first N-subjettiness axis \\
    \texttt{fj\_tau0\_trackEtaRel\_1} & Second smallest track $\Delta\eta$ relative to the jet axis, associated to the first N-subjettiness \\
             & \quad axis\\
    \texttt{fj\_tau0\_trackEtaRel\_2} & Third smallest track $\Delta\eta$ relative to the jet axis, associated to the first N-subjettiness axis\\
    \texttt{fj\_tau1\_trackEtaRel\_0} & Smallest track $\Delta\eta$ relative to the jet axis, associated to the second N-subjettiness axis\\
    \texttt{fj\_tau1\_trackEtaRel\_1} & Second smallest track $\Delta\eta$ relative to the jet axis, associated to the second N-subjettiness\\
             & \quad axis\\
    \texttt{fj\_tau1\_trackEtaRel\_2} & Third smallest track $\Delta\eta$ relative to the jet axis, associated to the second N-subjettiness  \\
             & \quad axis\\
    \texttt{fj\_tau\_flightDistance2dSig\_0} & Transverse (2D) flight distance significance between the PV and the SV with the smallest \\
             & \quad uncertainty on the 3D flight distance associated to the first N-subjettiness axis \\
    \texttt{fj\_tau\_flightDistance2dSig\_1} & Transverse (2D) flight distance significance between the PV and the SV with the smallest \\
             & \quad uncertainty on the 3D flight distance associated to the second N-subjettiness axis \\
    \texttt{fj\_tau\_vertexDeltaR\_0} & $\Delta R$ between the first N-subjettiness axis and SV direction \\
    \texttt{fj\_tau\_vertexEnergyRatio\_0} & SV energy ratio for the first N-subjettiness axis, defined as the total energy of all SVs \\
             & \quad associated with the first N-subjettiness axis divided by the total energy of all the \\
             &  \quad tracks associated with the AK8 jet that are consistent with the PV \\
    \texttt{fj\_tau\_vertexEnergyRatio\_1} & SV energy ratio for the second N-subjettiness axis\\
    \texttt{fj\_tau\_vertexMass\_0} & SV mass for the first N-subjettiness axis, defined as the invariant mass of all tracks from  \\
             & \quad SVs associated with the first N-subjettiness axis \\
    \texttt{fj\_tau\_vertexMass\_1} & SV mass for the second N-subjettiness axis \\
    \texttt{fj\_trackSip2dSigAboveBottom\_0} & Track 2D signed impact parameter significance of the first track lifting the combined  \\
             & \quad invariant mass of the tracks above the \PQb hadron threshold mass (5.2 \GeV) \\
    \texttt{fj\_trackSip2dSigAboveBottom\_1} & Track 2D signed impact parameter significance of the second track lifting the combined \\
             & \quad invariant mass of the tracks above the \PQb hadron threshold mass (5.2 \GeV) \\
    \texttt{fj\_trackSip2dSigAboveCharm\_0} & Track 2D signed impact parameter significance of the first track lifting the combined  \\
             & \quad invariant mass of the tracks above the \PQc hadron threshold mass (1.5 \GeV) \\
    \texttt{fj\_trackSipdSig\_0} & Largest track 3D signed impact parameter significance\\
    \texttt{fj\_trackSipdSig\_1} & Second largest track 3D signed impact parameter significance \\
    \texttt{fj\_trackSipdSig\_2} & Third largest track 3D signed impact parameter significance \\
    \texttt{fj\_trackSipdSig\_3} & Fourth largest track 3D signed impact parameter significance \\
    \texttt{fj\_trackSipdSig\_0\_0} & Largest track 3D signed impact parameter significance associated to the first \\
             & \quad N-subjettiness axis \\
    \texttt{fj\_trackSipdSig\_0\_1} & Second largest track 3D signed impact parameter significance associated to the first \\
             & \quad N-subjettiness axis \\
    \texttt{fj\_trackSipdSig\_1\_0} & Largest track 3D signed impact parameter significance associated to the second \\
             & \quad N-subjettiness axis\\
    \texttt{fj\_trackSipdSig\_1\_1} & Second largest track 3D signed impact parameter significance associated to the second \\
             & \quad N-subjettiness axis\\
    \texttt{fj\_z\_ratio} & $z$ ratio variable as defined in Ref.~\cite{Sirunyan:2017ezt}
  \end{tabular} 
\end{ruledtabular} }
  \label{tab:jet_features}
\end{table*}

\newlength\partTableWidth
\ifpreprint{\setlength\partTableWidth{0.95\textwidth}}{\setlength\partTableWidth{\textwidth}}

\begin{table*}[htpb!]
    \centering
    \caption{Additional features for charged or neutral particles. The all-particle IN model uses these features.}
    \resizebox{\partTableWidth}{!}{
\begin{ruledtabular}
    \begin{tabular}{ll}
      Variable & Description\\\hline
      \texttt{pfcand\_ptrel} & $\pt$ of the charged or neutral particle divided by the $\pt$ of the AK8 jet \\
      \texttt{pfcand\_erel} & Energy of the charged or neutral particle divided by the energy of the AK8 jet \\
      \texttt{pfcand\_phirel} & $\Delta\phi$ between the charged or neutral particle and the AK8 jet axis \\
      \texttt{pfcand\_etarel} & $\Delta\eta$ between the charged or neutral particle and the AK8 jet axis \\
      \texttt{pfcand\_deltaR} & $\Delta R$ between the charged or neutral particle and the AK8 jet axis \\
      \texttt{pfcand\_puppiw} & Pileup per particle identification (PUPPI) weight~\cite{PUPPI} for the charged or neutral particle \\
      \texttt{pfcand\_drminsv} & $\Delta R$ between the associated SVs and the charged or netural particle \\
      \texttt{pfcand\_drsubjet1} & $\Delta R$ between the charged or neutral particle and the first soft drop subjet \\
      \texttt{pfcand\_drsubjet2} & $\Delta R$ between the charged or neutral particle and the second soft drop subjet \\
      \texttt{pfcand\_hcalFrac} & Fraction of energy of the charged or neutral particle deposited in the hadron calorimeter\\
    \end{tabular} 
\end{ruledtabular}}
\label{tab:part_features}
\end{table*}

\clearpage

\bibliography{bib}

\end{document}